\documentclass[aps,prd,final,singlecolumn,superscriptaddress,
floatfix,preprintnumbers,nofootinbib,showpacs,showkeys]{revtex4}
\usepackage{dcolumn} 
\usepackage{amssymb}
\usepackage{amsbsy}
\usepackage[tbtags]{amsmath}
\usepackage{mathtools}

\usepackage{graphicx}

\usepackage{floatflt}
\usepackage[caption=false]{subfig}
\usepackage{overpic}
\usepackage{placeins} % defines the \FloatBarrier

\usepackage{supertabular}

\usepackage{xcolor}
\usepackage{ulem}
\usepackage{wrapfig}

\usepackage{xspace}
\usepackage{slashed}

\usepackage{hyperref}  
\usepackage{verbatim}

\usepackage{dsfont}

\newcommand{\Eins}{\mathds{1}}

\def\clap#1{\hbox to 0pt{\hss#1\hss}}

\newcommand{\bc}{\begin{center}}
\newcommand{\ec}{\end{center}}
\def\t0{t\text{$=$}0}
\def\ix0{\xi\text{$=$}0}
\def\wg{{g_{1T}}}

\newcommand{\bea}{\vspace{-0mm}\begin{eqnarray}}
\newcommand{\eea}{\end{eqnarray}}

\newcommand{\units}[1]{\ensuremath{\,\mathrm{#1}}}
\newcommand{\bra}[1]{\left\langle #1 \right|}
\newcommand{\ket}[1]{\left| #1 \right\rangle}

\newcommand{\prp}{\perp}

\newcommand{\GammaOp}{\Gamma}

\newcommand{\lat}{{\text{lat}}}

\newcommand{\tcdot}{{\cdot}}

\newcommand{\Wline}[1]{\ensuremath{{\mathcal{U}}[#1]}}
\newcommand{\WlineC}[1]{\ensuremath{{\mathcal{U}}{[#1]}}}

\newcommand{\vect}[1]{\ensuremath{\boldsymbol{#1}}}
\newcommand{\vprp}[1]{\vect{#1}_{\mathrm T}}

\newcommand{\quark}{q}

\newcommand{\tAmp}{\widetilde{A}}
\newcommand{\tBmp}{\widetilde{B}}

\newcommand{\nplus}{\ensuremath{\bar n}}
\newcommand{\nminus}{\ensuremath{n}}

\newcommand{\TMD}{TMD\xspace}
\newcommand{\TMDs}{TMDs\xspace}

\newcommand{\toddmark}[1]{\ensuremath{\Bigg[#1\Bigg]_{\text{\tiny{odd}}}}}

\newcommand{\fourint}{\ensuremath{\int_\mathcal{F}}}

\newcommand{\myeps}{\ensuremath{\epsilon}}

\newcommand{\bvec}{b}
\newcommand{\softf}{\mathcal{S}}
\newcommand{\mN}{m_N}
\newcommand{\nf}{n_f}

\newcommand{\MSbar}{\overline{\text{MS}}}

\newcommand{\kei}{k}

\newcommand{\unsub}{\text{unsubtr.}}
\newcommand{\zetahat}{{\hat \zeta}}

\begin{document}  

\title{Sivers and Boer-Mulders observables from lattice QCD}

\preprint{MIT-CTP~4324, JLAB-THY-11-1462}

\author{B.U.~Musch}
  \affiliation{Theory Center, Jefferson Lab, Newport News, VA 23606, USA}
   \email{bmusch@jlab.org}
\author{Ph.~H\"agler}
  \affiliation{Institut f\"ur Kernphysik, Johannes Gutenberg-Universit\"at Mainz, D-55128 Mainz, Germany}
  \email{haegler@p2h.de}
\author{M.~Engelhardt}
  \affiliation{Department of Physics, New Mexico State University, Las Cruces, NM 88003-8001, USA}
  \email{engel@nmsu.edu}
\author{J.W.~Negele}
  \affiliation{Center for Theoretical Physics, Massachusetts Institute of Technology, Cambridge, Massachusetts 02139, USA}
\author{A.~Sch\"afer}
  \affiliation{Institut f\"ur Theoretische Physik, Universit\"at
  Regensburg, 93040 Regensburg, Germany}

\date{\today}

\begin{abstract} 

We present a first calculation of transverse momentum dependent nucleon
observables in dynamical lattice QCD employing non-local operators with
staple-shaped, ``process-dependent" Wilson lines. The use of staple-shaped
Wilson lines allows us to link lattice simulations to TMD effects
determined from experiment, and in particular to access non-universal,
naively time-reversal odd TMD observables. We present and discuss results
for the generalized Sivers and Boer-Mulders transverse momentum shifts
for the SIDIS and DY cases. The effect of staple-shaped Wilson lines
on T-even observables is studied for the generalized tensor charge and a
generalized transverse shift related to the worm gear function $g_{1T}$.
We emphasize the dependence of these observables on the staple extent
and the Collins-Soper evolution parameter. Our numerical calculations
use an $\nf = 2{+}1$ mixed action scheme with domain wall valence fermions
on an Asqtad sea and pion masses $369 \units{MeV}$ as well as
$518 \units{MeV}$.
\end{abstract}



\maketitle

\section{Introduction}
\label{sec-intro}
The picture of the nucleon as a system of interacting quarks and gluons naturally leads to the question about the intrinsic motion of these elementary particles inside the proton or neutron. 
This intrinsic motion, specifically with respect to the transverse momentum, 
can be described in terms of Transverse Momentum Dependent Parton Distribution Functions (\TMDs), see, e.g., 
chapter 2 of Ref. \cite{Boer:2011fh} for a recent review.
TMDs for quarks, generically denoted by $f_1(x,\vprp{k}^2)$, $g_1(x,\vprp{k}^2)$, etc., encode essential information about the distribution of partons with respect to the longitudinal momentum fraction, $x$, and intrinsic quark transverse momentum, $\vprp{k}$.
With certain restrictions in mind, they have an intuitively appealing interpretation as three-dimensional probability densities \cite{Bacchetta:1999kz,Collins:2003fm}. 
TMDs can, for example, be studied on the basis of angular asymmetries observed in processes such as Semi Inclusive Deep Inelastic Scattering (SIDIS) using suitable QCD factorization theorems that go beyond the standard collinear factorization, 
see, e.g., Refs.~\cite{Ji:2004wu,Ji:2004xq,Aybat:2011zv,CollinsBook2011}.
In contrast to the usual collinear PDFs, TMDs turn out to be in general non-universal, i.e., process-dependent.
The process dependence arises from the difference in the final and initial state interactions in SIDIS and Drell-Yan scattering, respectively.
On the theoretical level, it can be understood as an intriguing consequence of the local color gauge invariance of the strong interaction
and the corresponding non-trivial gauge-link structures.
Specifically, QCD factorization leads to the remarkable prediction that the naively time reversal odd (T-odd) TMDs, in particular
the Sivers and Boer-Mulders functions, differ in sign for DY compared to SIDIS, $f^{\text{T-odd,SIDIS}}=-f^{\text{T-odd,DY}}$.
The implications and consequences of these observations continue to stir intense interest of many theoreticians and experimentalists,
as a number of fundamental questions and interesting puzzles remain to be addressed.
Motivated by promising experimental results from COMPASS, HERMES and JLab 
(see, e.g., \cite{COMPASS:2008dn,HERMES:2009ti,Avakian:2010ae} and references therein), 
as well as considerable progress on the theoretical and phenomenological sides during recent years, 
an essential part of the physics program of future facilities will therefore be targeted in this direction, 
including JLab $12\units{GeV}$ and the proposed EIC at JLab or BNL. 
 
Theoretical calculations of \TMDs from first principles require non-perturbative methods such as lattice QCD. 
In previous works, we have introduced and explored techniques that allow the computation of the underlying amplitudes
on the lattice using non-local operators \cite{MuschThesis2,Hagler:2009mb,Musch:2010ka}. 
Our numerical studies for a ``process-independent", direct gauge link geometry already produced encouraging results.
In this work, we present a first exploratory lattice study employing a more complex, ``process-dependent" link geometry 
that gives us rather direct access to highly interesting T-odd observables. 

In section \ref{sec-formalism}, we present the formalism and techniques
required for our calculations,
and provide definitions of the relevant T-odd and T-even TMD observables.
After a short introduction to the lattice computations at the beginning of section \ref{sec-latticecalc}, we continue with a presentation 
and discussion of our numerical results for the generalized shifts and tensor charge.
A summary and conclusions are given in section \ref{sec-summary}.

\section{Formalism}
\label{sec-formalism}

\subsection{Definition of \TMDs} 
\label{sec-TMDdef}

In a relativistic quantum field theory, the question ``What is the probability to find a quark with a given momentum $\kei$ inside the proton?'' needs to be stated more precisely. 
First of all, it turns out to be advantageous to formulate everything in light cone coordinates, see appendix \ref{sec-conv}, 
and to consider a frame of reference where the nucleon has large momentum in $z$-direction, i.e., $P^+ \gg m_N$, $\vprp{P} = 0$.
In light cone coordinates, the components $\kei^+$,~$\vprp{\kei}$,~$\kei^-$ of the quark momentum $\kei$ scale as $P^+/m_N$,~$1$,~$m_N/P^+$, respectively, under boosts along the $z$-axis. Thus the longitudinal momentum fraction of the quark $x = k^+ / P^+$ and its transverse momentum $\vprp{\kei}$ are invariant under boosts along the $z$-axis, while the $k^-$ component is suppressed. 
This leads to the concept of transverse momentum dependent parton distribution functions (\TMDs), which are functions of the longitudinal momentum fraction $x \equiv \kei^+ / P^+$ and of the quark transverse momentum $\vprp{\kei}$.
The transverse momentum components $\vprp{\kei}$ are particularly interesting, because they describe an intrinsic motion of the quarks inside the proton that occurs independent of the momentum of the proton itself. This gives us a unique picture of the dynamics inside the proton. Moreover, the \TMDs are an important ingredient in our understanding of the origin of large angular- and spin-asymmetries found in experiments studying, e.g., semi-inclusive deep inelastic scattering (SIDIS) or the Drell-Yan process (DY).

In a naive approach based on a theory quantized on the light front, one obtains a momentum dependent number density of quarks from $f_1(x,\vprp{k}) \sim \frac{1}{2} \sum_{\Lambda = \pm 1} \sum_{\lambda = \pm 1}   |a_{\lambda,\quark}(x,\vprp{\kei}) \ket{P,S}|^2$ (up to normalization factors), where $a_{\lambda,\quark}$ is an annihilation operator of quarks of flavor $\quark$ and helicity $\lambda$. The average over nucleon helicities $\frac{1}{2}\sum_{\Lambda \pm 1}$ implements an average over the spin $S$ in the nucleon state $\ket{P,S}$. 
In this example, the TMD $f_1(x,\vprp{k})$ describes the distribution of unpolarized quarks in an unpolarized nucleon.
Rewriting the annihilation operator in terms of local quark field operators
$\bar q$ and $q$ reveals a problem: $f_1(x,\vprp{k})$ is a Fourier transform
of the matrix element $\bra{P,S} \bar q(0) \gamma^+ q(\bvec) \ket{P,S}$ with
respect to the position $\bvec$, and the bi-local operator
$\bar q(0) \gamma^+ q(\bvec)$ is not gauge invariant, see
Ref. \cite{Collins:2003fm} for a review of the issue. Gauge invariance
can be restored by inserting a Wilson line $\WlineC{\mathcal{C}_\bvec}$
between the quark fields, as defined in appendix \ref{sec-conv}. The Wilson
line introduces divergences that cannot be treated by conventional dimensional
regularization \cite{Collins:2008ht}. Several different schemes have been
proposed in the literature as to how to subtract those divergences
\cite{Collins:2003fm,Collins:2004nx,Ji:2004wu,Hautmann:2007uw,Chay:2007ty,Cherednikov:2008ua,Collins:2008ht,Aybat:2011zv,CollinsBook2011,Aybat:2011ge},
see Ref. \cite{Cherednikov:2011ft} for a recent comparison. In general,
these schemes require the introduction of a so-called soft factor
$\tilde \softf$ inside the defining correlator of \TMDs. The starting point
for our discussion of \TMDs is thus a correlator of the general form
\begin{align}
	\Phi^{[\Gamma]}(\kei,P,S;\ldots)\ \equiv\ \ 
	\int \frac{d^4 \bvec}{(2\pi)^4}\, e^{i \kei\cdot \bvec}\,
		\frac{\overbrace{\rule{0em}{1.2em}
		\frac{1}{2}\, \bra{P,S}\ \bar{q}(0)\, \Gamma\ \WlineC{\mathcal{C}_\bvec}\ q(\bvec)\ \ket{P,S}
		}^{\displaystyle \equiv \widetilde \Phi_\unsub^{[\Gamma]}(\bvec,P,S;\ldots)}}
		{ \widetilde{\softf}(\bvec^2;\ldots) }
		%{ \underbrace{ \widetilde{\mathcal{S}}(\elll^2,\ldots) }_{\displaystyle \text{soft factor}}}
		\label{eq-corr}	
\end{align}
The detailed properties of the Wilson line $\WlineC{\mathcal{C}_\bvec}$ and the soft factor need to be specified by additional parameters, which we indicate by the dots ``$\ldots$'' for now and which will be discussed later. Moreover,
all objects above implicitly depend on a UV renormalization scale $\mu$.

In Eq.~\eqref{eq-corr}, $\tilde \softf$ stands somewhat symbolically for an
expression that can, depending on the formalism, involve several vacuum
expectation values.
% For example, in Ref. \cite{Collins}, our  $\widetilde{\softf}(\bvec^2;\ldots)$ corresponds to the expression
% \begin{equation}
%   \widetilde{\softf}(\bvec^2;\ldots) = \sqrt{ 
%     \frac{ 
%       \bra{0} \Wline{\bvec_\prp / 2, \bvec_\prp / 2 + \infty \eta v} \ket{0}
%       }{ \bra{0} \Wline{\bvec_\prp / 2, \bvec_\prp / 2 + \infty \eta v} \ket{0} }
%     }
% \end{equation}
For example, in the scheme developed in Refs. \cite{Aybat:2011zv,CollinsBook2011,Aybat:2011ge}, our factor $\widetilde{\softf}(\bvec^2;\ldots)$ would be (using the notation of those references) 
\begin{equation}
\widetilde{\softf}(\bvec^2;\ldots) = 
  \sqrt{ \frac
    { \tilde S_{(0)}( \vprp{\bvec}, + \infty, -\infty )\ \tilde S_{(0)}( \vprp{\bvec}, y_s, -\infty ) }
    { \tilde S_{(0)}( \vprp{\bvec}, + \infty, y_s ) }
  }
\end{equation}
where each of the objects $\tilde S_{(0)}( \vprp{\bvec}, \ldots )$ is a
vacuum expectation value of Wilson line structures.
In this specific framework, the starting point of the discussion is space-like Wilson lines. Some Wilson lines remain tilted away from the light cone, leading to the dependence on the rapidity parameter $y_s$ in the above expression. Other Wilson lines, including those contained in $\WlineC{\mathcal{C}_\bvec}$ in the numerator of Eq.~\eqref{eq-corr}, are brought back to the light cone in the sense of a limit, as indicated symbolically by $+\infty$ and $-\infty$ in the equation above, see Refs. \cite{Aybat:2011zv,CollinsBook2011} for details. As we will see in section \ref{sec-linkgeom} and below, certain matrix elements with space-like structures of Wilson lines are directly accessible on the Euclidean lattice, whereas taking the light-cone limit is only possible in the form of a numeric limit and technically challenging.
For the purposes of our treatment, however, we do not need to go into any
detail concerning the definition of $\widetilde{\softf}(\bvec^2;\ldots)$
in any particular framework, 
%For the purposes of our treatment, $\tilde \softf$ does not need to be
%specified more explicitly,
since it will cancel in the observables we consider. 

Integrating the correlator over the suppressed momentum
component $k^-$ yields 
\begin{align}
	\Phi^{[\Gamma]}(x,\vprp{k};P,S;\ldots) & \equiv \int dk^{-} \Phi^{[\Gamma]}(\kei,P,S;\ldots) \nonumber \\ 
	& =\int \frac{d^2 \vprp{\bvec}}{(2\pi)^2} \int \frac{d (\bvec \tcdot P)}{(2\pi) P^+}\ e^{i x (\bvec \tcdot P) - i\vprp{\bvec} \tcdot \vprp{\kei}}\,
		\left. \frac{\frac{1}{2}\, \bra{P,S}\ \bar{q}(0)\, \Gamma\ \WlineC{\mathcal{C}_\bvec}\ q(\bvec)\ \ket{P,S}}
		{ \widetilde{\mathcal{S}}(-\vprp{\bvec}^2;\ldots) } \right|_{b^+ = 0} \ .
\end{align}
Notice that integrating over $\kei^-$ corresponds to setting $b^+=0$. As a consequence, $x \leftrightarrow (\bvec \tcdot P)$ and $\vprp{k} \leftrightarrow \vprp{b}$ act as independent pairs of Fourier conjugate variables in the expression above. The above correlator can be decomposed into \TMDs.
For choices of the Dirac matrix $\Gamma$ that project onto leading twist, one obtains \cite{Ralston:1979ys,Tangerman:1994eh,Mulders:1995dh,Goeke:2005hb}
\begin{align}
	\Phi^{[\gamma^+]}(x,\vprp{k};P,S,\ldots) & = f_1 - \toddmark{\frac{\myeps_{ij}\, \vect{k}_{i}\, \vect{S}_{j}}{m_N}\ f_{1T}^\prp} \label{eq-phigammaplus}\ , \displaybreak[0]  \\ 
	\Phi^{[\gamma^+\gamma^5]}(x,\vprp{k};P,S,\ldots) & = \Lambda\, g_{1} + \frac{\vprp{k} \cdot \vprp{S}}{m_N}\ g_{1T} \label{eq-phigammaplusgfive} \ , \displaybreak[0] \\
	\Phi^{[i\sigma^{i+}\gamma^5]}(x,\vprp{k};P,S,\ldots) & = \vect{S}_i\ h_{1}   +  \frac{(2 \vect{k}_i \vect{k}_j - \vprp{k}^2 \delta_{ij}) \vect{S}_j}{2 m_N^2}\, h_{1T}^\prp + \frac{\Lambda \vect{k}_i}{m_N} h_{1L}^\prp +   \toddmark{\frac{\myeps_{ij} \vect{k}_{j}}{m_N} h_1^\prp}\ .\label{eq-phisigmaplusi} 
\end{align}
The \TMDs $f_1$, $g_1$, $h_1$, $g_{1T}$, $h_{1L}^\perp$, $h_{1T}^\perp$, $f_{1T}^\perp$ and $h_1^\perp$ are functions of $x$, $\vprp{\kei}^2$, $\mu$ and further parameters related to regularization and link geometry.
The structures shown in brackets $[\ ]_\text{odd}$ involve so-called naively time reversal odd (T-odd) \TMDs, namely the Sivers function $f_{1T}^\perp$ \cite{Sivers:1989cc} and the Boer-Mulders function $h_1^\perp$ \cite{Boer:1997nt}. The origin of the above parametrization and the special role of T-odd \TMDs will become clear after we have discussed the geometry of the gauge link path
$\mathcal{C}_\bvec $ and symmetry transformation properties.

\subsection{General strategy}

At this point, several remarks are in order as to how we aim to introduce \TMD observables that can be accessed with lattice QCD using a non-local operator technique.
Due to the underlying operator structure, the situation is quite different from that of standard collinear
PDFs and offers unique opportunities and challenges. 

As an introductory example, consider the definition of a standard PDF in the unpolarized case,
\begin{equation*}
  f_1(x) \equiv \frac{1}{2 (2\pi)} \int db^- e^{i x P^+ b^-} \, \bra{P,S}\ \bar{q}(0)\, \gamma^+\ \Wline{0,n b^-}\ q(n b^-)\ \ket{P,S} \ .
\end{equation*}
%Here $\ket{P,S}$ represents a proton state with momentum $P$ and spin vector $S$, and $\gamma^+ = ( \gamma^0 + \gamma^3) / \sqrt{2}$ is a Dirac matrix. The quark field operators $\bar{q}(0)$ and $q(n b^-)$ are separated along a light-like direction $n$. 
For PDFs, the gauge link $\Wline{0,n b^-}$ is simply a straight, light-like Wilson line of finite extent connecting the two quark field operators \cite{Collins:1981uw}. No continuous Lorentz transformation exists that allows us to ``rotate'' the non-local operator $\bar{q}(0)\, \gamma^+\ \Wline{0,n b^-}\ q(n b^-)$ into Euclidean space. The light-like separation stays always light-like, but in Euclidean space objects cannot have any extent in (Minkowski-) time. 
%To avoid this problem, the standard approach of calculating PDFs on the lattice is based on an expansion in terms of \emph{local} operators.
As a consequence, one is forced
to invoke the operator product expansion to cast the calculation in
terms of local matrix elements which can be accessed using lattice QCD.

The situation for \TMDs differs fundamentally in several aspects:
\begin{enumerate}
  \item The separation $b$ of the quark field operators has an additional transverse component, $\bvec = n \bvec^- + \bvec_\perp$. Thus, in general, this separation is space-like. This opens the possibility of a direct representation of the non-local operator in Euclidean space.
  \item The geometry of the gauge link $\WlineC{\mathcal{C}_\bvec}$ is more complicated, depends to a certain degree on the experiment under consideration and in general extends out to infinity. As a result, it becomes questionable whether an expansion in terms of local operators is possible at all. 
  \item Regularization is more complicated, leading to the introduction of the soft factor $\tilde \softf$ and additional regularization parameters beyond the usual renormalization scale $\mu$ of the $\overline{\mathrm{MS}}$ scheme. 
%The usual dimensional regularization procedure is not sufficient to eliminate the divergences of the gauge link, see, e.g. Ref. \cite{Cherednikov}. Various schemes have been developed that address those divergences, see, e.g., Refs. \cite{bb}. In general, these schemes introduce a subtraction factor (or, soft factor) in the definition of a \TMD and  additional regularization parameters beyond the usual $\overline{\mathrm{MS}}$ renormalization scale. For some schemes evolution equations in those additional parameters are known, see, e.g, \cite{bb}. Note however, that for most schemes, the naive connection between \TMDs $f(x,\vprp{k})$ and PDFs $f(x)$ through $\vprp{k}$-integration, $f(x) = \int d\vprp{k} f(x,\vprp{k})$, is not valid \cite{Cheredikov}.     
\end{enumerate}

The first two items listed above are our main motivation to develop a technique for lattice studies of \TMDs based on non-local operators.
It should be emphasized that this technique can only work for the analysis of certain \TMD-related observables within a limited kinematical range. The method cannot be applied to study the $x$-dependence of PDFs directly, without the use of non-trivial extrapolations.

In previous publications \cite{Hagler:2009mb,Musch:2010ka}, it was
demonstrated that the non-local operator technique is quite promising and
produces interesting results, for a simplified gauge link geometry, at
least on a qualitative level. The crucial connection between the formalism
in Minkowski space and the results from Euclidean space is provided through
a parametrization in terms of invariant amplitudes\footnote{Note that the
symbol $l$ in Ref. \cite{Hagler:2009mb,Musch:2010ka} corresponds to
$-\bvec$ in the present study.} $\tAmp_i(\bvec^2,\bvec \tcdot P)$. By
virtue of their Lorentz-invariance, the calculation of these amplitudes
can be performed in any desired Lorentz frame. In particular, for the
generic off-light cone kinematics appropriate for \TMDs, there is no
obstacle to performing the calculation in a frame in which the nonlocal
operator in question is defined entirely at one fixed time. In this frame,
one can cast the computation of the nonlocal matrix element
in terms of a Euclidean path integral, evaluated employing the standard
methods of lattice QCD.

The study at hand builds directly on Ref. \cite{Musch:2010ka}, and we refer the reader to that publication for an introduction to the essential principles of the methology. One of the remaining challenges identified in Ref. \cite{Musch:2010ka} concerns the geometry of the gauge link. In the present study, we replace the simple straight connection by a staple-like path that corresponds more accurately to the situation in phenomenology. We stress that these gauge link
structures are part of the established phenomenological framework, which
we take as given, and not a new assumption related to our use of
lattice QCD as a calculational method. Whereas our results depend on
the gauge link structure, specific physical processes such as SIDIS and
DY unambiguously correspond to definite instances of that structure.
Throughout our discussion, we clearly identify the SIDIS and DY limits
of our data.

It is important to point out that our assumptions about the operator
structure of \TMDs rely on factorization arguments that are much more
involved than for the usual PDFs. In fact, one must be judicious
concerning the classes of reactions for which it can be assumed that
a factorization framework with well-defined \TMDs exists. For example,
it has been realized recently \cite{Rogers:2010dm,Aybat:2011vb} 
that TMD factorization generally fails for large reaction classes,
in particular processes with multiple hadrons in both the initial and
the final state. While the consequences of this observation are not yet
all known, it appears certain that to develop a TMD framework for these
processes, a fundamental change of perturbative QCD techniques is needed.
It could, e.g., very well be that measurement-independent cross-sections
are simply not defined for certain reaction classes and that, instead,
the appropriate quantities will be entanglement amplitudes which then
have to be folded with quantities encoding the measurement 
process \cite{RogersPriv}. In contradistinction to the aforementioned
classes of reactions, for other types of processes such as SIDIS and DY,
recent progress \cite{Aybat:2011zv,CollinsBook2011,Aybat:2011ge} indicates
that a valid definition of \TMDs based on factorization arguments indeed is
possible, within a scheme regularized employing space-like links. Promising
steps have been taken to develop the predictive capabilities of this
framework \cite{Aybat:2011ta}. A pertinent discussion is given in
section I of our previous publication \cite{Musch:2010ka}, with further
details to be found in the references therein and recent overviews in
Refs.~\cite{Cherednikov:2011ft,Aybat:2011vb}.
The point which we wish to emphasize here is that it is not the purpose
of our present work to critique or justify the various approaches to
defining \TMDs in terms of operators and matrix elements which have been
advanced in response to issues of factorization and regularization.
Instead, we will assume
that a good definition of \TMDs with a connection to phenomenology through
a valid factorization argument exists for certain classes of processes
such as SIDIS and DY; we focus exclusively on those \TMDs and do not aim
to contribute to discussions of factorization, fragmentation functions,
or related matters. Our starting point thus is the definition of \TMDs in
terms of a \TMD correlator of the rather general form \eqref{eq-corr}.

Working from this definition, we will moreover restrict ourselves to
observables in which the soft factor cancels, so that specifics of the
soft factor are not relevant for our results. Regarding the
necessary regularization of the gauge link, we pick the proposal that is
most suitable for our purposes, namely tilting the gauge link slightly
away from the light cone \cite{Collins:1981uk}, in a space-like direction
\cite{Aybat:2011zv,CollinsBook2011}. We stress that, in choosing this
approach to defining and regularizing \TMDs, we are led to consider kinematics
off the light cone from the very beginning, which makes a connection to
Euclidean lattice QCD feasible, as already noted further above. We emphasize
that, within this work, we do not aim to arrive at any statements concerning
the formal nature of the light-cone limit. We will, however, focus particularly
on the behavior of our numerical results as we extend the kinematic
region as far towards the light cone as possible.

One necessary step of a lattice calculation is to discretize the operators.
The discretization of non-local operators as we encounter them here is still
a rather new concept. An important assumption we make is that non-local
lattice operators composed of structures much larger than the lattice
spacing essentially renormalize in the same fashion as their counterparts
in the continuum, except that the renormalization parameters are specific
to the lattice action and the discretization prescription. We have given
reasons for this assumption and explored it numerically in
Ref.~\cite{Musch:2010ka}, see in particular sections III~D, IV~B, IV~C and
appendices B, D, G and H therein. However, we point out that a more rigorous
treatment would still be desirable. Especially the question of mixing
properties as one attempts to make contact with the local operator
formalism remains a challenge for the future.

Keeping the above remarks in mind, it is worthwhile summarizing the logic
underlying our treatment succinctly before laying out the details further
below:
\begin{enumerate}
\item
We start from a definition of \TMDs in terms of the correlator \eqref{eq-corr},
considering generic off-light cone kinematics from the very beginning.
\item
The correlator \eqref{eq-corr} is parametrized in terms of Lorentz-invariant
amplitudes, cf.~section~\ref{sec-parametr} below. This crucial step permits
one to transform results into different Lorentz frames in a simple manner.
\item
On this basis, we choose the Lorentz frame in which the nonlocal operator
entering \eqref{eq-corr} is defined at one single time as the one most
suitable for our calculation. We stress again that there is no obstacle
to this choice, since the separations in the operator are all space-like.
\item
In the aforementioned frame, the computation of the nonlocal matrix element
can be cast in terms of a Euclidean path integral and performed employing
the standard methods of lattice QCD.
\item
We form appropriate ratios of the extracted invariant amplitudes in which
soft factors and multiplicative renormalization factors cancel, such as
the $\vprp{k}$-shifts discussed in section~\ref{sec-latquan}, which, in
principle, represent measurable quantities. We particularly study the
approach to the SIDIS and DY limits in these quantities.
\end{enumerate}

\subsection{\TMDs in Fourier space and $x$-integration}

In essence, the lattice method we use allows us to evaluate the $b$-dependent matrix elements $\widetilde \Phi_\unsub^{[\Gamma]}(\bvec,P,S;\ldots)$ introduced in Eq. \eqref{eq-corr}. As a result, it is more direct and natural to state our results in terms of Fourier-transformed, $\vprp{b}$-dependent \TMDs and their $\vprp{b}$-derivatives. For a generic \TMD $f$ we define
\begin{align}
	\tilde f(x, \vprp{\bvec}^2;\ldots) & \equiv \int  d^2 \vprp{\kei} \, e^{i \vprp{\bvec} \cdot \vprp{\kei} }\; f(x, \vprp{\kei}^2;\ldots)
		= 2\pi \int      d|\vprp{\kei}| |\vprp{\kei}|\ J_0(|\vprp{\bvec}||\vprp{\kei}|)\  f(x, {\vprp{\kei}^2};\ldots )\ , \label{eq-FTMD} \\
	\tilde f^{(n)}(x, \vprp{\bvec}^2\ldots ) & \equiv n!\left( -\frac{2}{\mN^2}\partial_{\vprp{\bvec}^2} \right)^n \ 
	\tilde f(x, \vprp{\bvec}^2;\ldots ) 
		= \frac{ 2\pi \ n!}{(\mN^2)^n} \int  d |\vprp{\kei}| |\vprp{\kei}|\left( \frac{|\vprp{\kei}|}{|\vprp{\bvec}|}\right)^n J_n(|\vprp{\bvec}||\vprp{\kei}|)\  f(x, {\vprp{\kei}^2};\ldots )\ , \label{eq-dFTMD}
\end{align}
where the $J_n$ are Bessel functions of the first kind, and $\mN$ is the mass of the target hadron.
These objects and their potential phenomenological relevance have been discussed in detail in Ref. \cite{Boer:2011xd}. Moreover, evolution equations are naturally expressed in terms of the $\tilde f^{(n)}$, compare, e.g., Ref.~\cite{Idilbi:2004vb}. In the limit $|\vprp{\bvec}| \rightarrow 0$, one recovers conventional $\vprp{\kei}$-moments of \TMDs:
\begin{align}
	\tilde f^{(n)}(x, 0;\ldots ) & = \int  d^2 \vprp{\kei} \left( \frac{\vprp{\kei}^2}{2 \mN^2 }\right)^n  f(x, \vprp{\kei}^2;\ldots ) \equiv  f^{(n)}(x)\ .
	\label{eq-btder2}
\end{align}
However, it is known  \cite{Bacchetta:2008xw} that $\vprp{\kei}$-moments like $f_1^{(0)}(x)$ and $f_{1T}^{\perp(1)}(x)$ are ill-defined without further regularization. The problem is that the integral in the above equation diverges if the integrand does not fall off quickly enough in the region of large $\vprp{\kei}$, where the \TMDs $f(x,\vprp{\kei}^2)$ are perturbatively predictable. Even though $\vprp{\kei}$-moments may be more familiar to the reader, we therefore do not attempt to extrapolate to $\vprp{\bvec}=0$, but rather state our results at finite $|\vprp{\bvec}|$, where the $\vprp{\kei}$-integrals of Eqs. \eqref{eq-FTMD} and \eqref{eq-dFTMD} can be shown to be convergent in the relevant cases \cite{Boer:2011xd}. 

Information about the $x$-dependence of \TMDs can be obtained from the lattice via the Fourier-conjugate variable, $\bvec \tcdot P$ \cite{Hagler:2009mb,Musch:2010ka}.
However, the calculations performed in Euclidean space only allow us to access a limited range of $\bvec \tcdot P$, precluding us from performing a straightforward Fourier transform.
In this work, we limit ourselves to the study of $x$-integrated \TMDs
\begin{align}
	f^{[1]}(\vprp{\kei}^2;\ldots) \equiv \int_{-1}^1 dx\ f(x,\vprp{\kei}^2;\ldots)\ .
\end{align}
These are accessible from the data at $\bvec \tcdot P=0$.
Here, the superscript ${}^{[1]}$ denotes the first Mellin moment in $x$.
The integration is performed over the full range of $x$.
\TMDs evaluated at negative values of $x$ can be related to anti-quark distributions, see, e.g., \cite{Mulders:1995dh,Musch:2010ka} for details.

\subsection{Quantities suitable for lattice extraction}
\label{sec-latquan}

Certain ratios of $\vprp{\kei}$-moments of \TMDs have interesting physical interpretations. For example, consider 
\begin{align}
	\mN \frac{f_{1T}^{\perp(1)}(x)}{f_1^{(0)}(x)} = 
	\left. \frac{ \int d^2 \vprp{\kei}\, \vect{\kei}_y  \ \Phi^{[\gamma^+]}(x,\vprp{\kei},P,S;\ldots) }{ \int d^2 \vprp{\kei}  \phantom{\vect{\kei}_y}\ \Phi^{[\gamma^+]}(x,\vprp{\kei},P,S;\ldots) } \right|_{\displaystyle \vprp{S}=(1,0) }\, , 
\end{align}
where $\gamma^+$ projects on leading-twist.
In the context of the density interpretation of \TMDs mentioned in section \ref{sec-TMDdef}, the ratio above yields the average transverse momentum
in $y$-direction, for quarks with given longitudinal momentum fraction $x$ inside a proton polarized in $x$-direction. 
We will show below that quantities like this can be calculated rather directly on the lattice. 
For the reasons mentioned above, we limit ourselves to ratios formed from $x$-integrated quantities. Let us therefore consider 
\begin{align}
	\langle \vect{k}_y \rangle_{TU} \equiv \mN \frac{f_{1T}^{\perp[1](1)}}{f_1^{[1](0)}}  \ .
	\label{eq-sivshift}
\end{align}
Ignoring the role of anti-quarks, this ratio, called in the following ``Sivers shift'', represents the average transverse momentum of unpolarized (``U'') quarks  orthogonal to the transverse (``T'') spin of the nucleon.
Note, however, that the denominator $f_1^{[1](0)}$ arises from a difference of quarks and anti-quarks and thus 
gives the number of valence quarks in the nucleon.
On the other hand, in the numerator $f_{1T}^{\perp[1](1)}$, the average transverse momentum of quarks and anti-quarks is summed over \cite{Mulders:1995dh,Musch:2010ka}.
A profound interpretation of $f_{1T}^{\perp[1](1)}$ in impact parameter space has been given in Ref.~\cite{Burkardt:2003uw}. 
However, as mentioned before, understanding $f_{1T}^{\perp[1](1)}$ simply as a $\vprp{\kei}$-weighted \TMD is problematic, since the $\vprp{k}$-integral is expected to be UV divergent. 
A natural way of circumventing this divergence is to generalize the Sivers shift to an expression in terms of the Fourier-transformed \TMDs:
\begin{align}
	\langle \vect{k}_y \rangle_{TU}(\vprp{\bvec}^2;\ldots) \equiv \mN \frac{\tilde f_{1T}^{\perp[1](1)}(\vprp{\bvec}^2;\ldots)}{\tilde f_1^{[1](0)}(\vprp{\bvec}^2;\ldots)}  \ .
\end{align}
This is the type of quantity that we investigate in the present study.  In the limit $\vprp{\bvec}^2=0$ we recover the Sivers shift \eqref{eq-sivshift}, because the Fourier transformed \TMDs $\tilde f_{1T}^{\perp[1](1)}$ and $\tilde f_1^{[1](0)}$ coincide with the moments $f_{1T}^{\perp[1](1)}$ and $f_1^{[1](0)}$, respectively. We are, however, interested in the generalized Sivers shift for non-zero $\vprp{\bvec}^2$, where the said UV-divergence disappears. The variable $\vprp{b}^2$ effectively acts as a regulator. Moreover, the $\vprp{\bvec}$-dependence allows us to study differences in the widths of distributions on a qualitative level.

\subsection{Link geometry}
\label{sec-linkgeom}

The prescription for the geometry of the gauge link path $\mathcal{C}_\bvec$ affects both the number of allowed structures appearing in Eqs. \eqref{eq-phigammaplus}-\eqref{eq-phisigmaplusi} and the numerical result for the \TMDs. We therefore need to ask which link geometries are appropriate. 

\begin{figure}[btp]
	\centering%
	\subfloat[][]{%
		\label{fig-link-straight}%\
		\includegraphics[trim=0 0 120 0,clip=true]{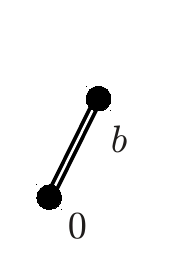}
		}\qquad%
	\subfloat[][]{%
		\label{fig-link-staple}%
		\includegraphics[]{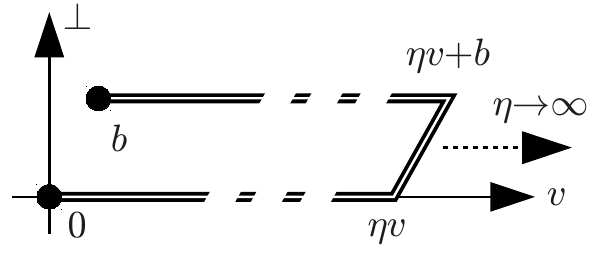}
		}%
	\caption[SIDIS diagram]{%
		\subref{fig-link-straight}\ %
			Straight gauge link.\ \ \par			
		\subref{fig-link-staple}\ %
			Staple-shaped gauge link as in SIDIS and DY.\par%
		\label{fig-links}%
		}
\end{figure}

The simplest link geometry is a straight line connecting the quark fields at $0$ and $\bvec$, see Fig. \ref{fig-link-straight}. \TMDs with straight gauge links have been studied on the lattice in Refs. \cite{Hagler:2009mb,Musch:2010ka}. While these ``process-independent'' \TMDs are interesting from a theoretical point of view in their own right, it is so far not known how to relate these quantitatively to the \TMDs that play a role in scattering experiments. The operator with straight gauge links offers the largest possible degree of symmetry. As a result, T-odd \TMDs vanish for straight gauge links.

For \TMDs that allow us to describe measurable effects in scattering experiments such as SIDIS or DY, the form of the gauge link is largely dictated by the physical process. To understand scattering experiments at high momentum transfer $Q$, one tries to apply approximations valid for large $Q$ that separate hard, perturbative and soft, non-perturbative scales in the dominant physical processes in order to arrive at an expression for the cross section in factorized form.
In the standard collinear approximation, all internal transverse momenta are integrated out and conventional parton distribution functions and fragmentation functions are used to describe the process. In certain kinematical regions this approximation is insufficient. An example is SIDIS, where the momentum $P_h$ of one of the final state hadrons is measured after a lepton-nucleon collision at large momentum transfer $Q$. The transverse momentum dependent formalism is needed when the transverse momentum component $\vect{P}_{h\perp}$ is small with respect to $Q$, see, e.g., Ref. \cite{Bacchetta:2008xw} for an in-depth discussion.

\begin{figure}[btp]
	\centering%
	\includegraphics[width=0.5\linewidth]{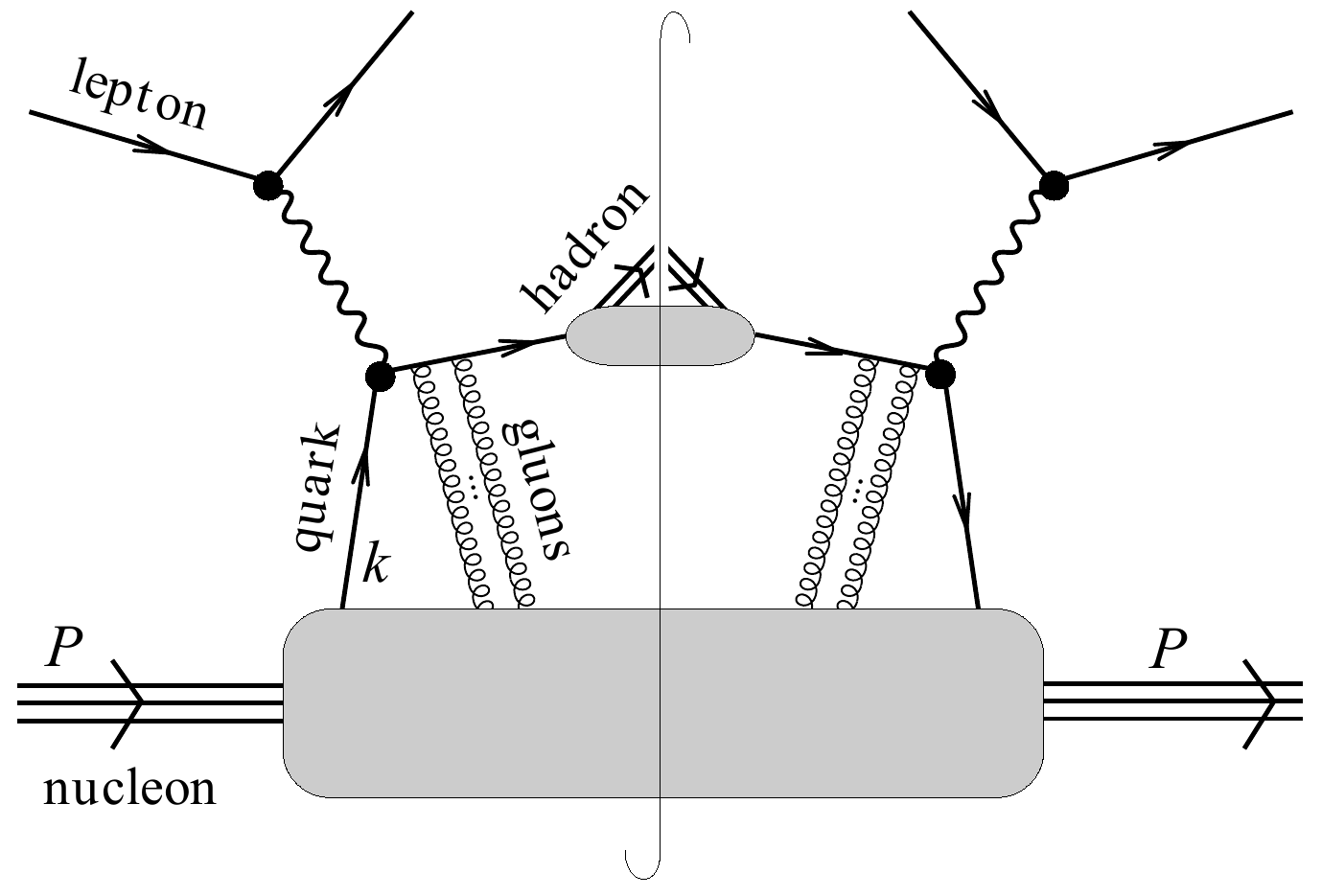}
	\caption{%
		Illustration of the leading contribution to SIDIS in factorized form.
		\label{fig-SIDISdiagram}
		}
\end{figure}

The leading diagram for SIDIS
is shown in a simplified, factorized form in Fig. \ref{fig-SIDISdiagram}. 
The lower shaded bubble in the diagram represents the structure parametrized by \TMDs. A gauge link in the \TMD correlator arises naturally as an idealized, effective, resummed description of the gluon exchanges between the ejected quark and the remainder of the nucleon in the evolving final state, see, e.g., Ref.~\cite{Pijlman:2006tq} for a review.
The gauge link roughly follows the direction of the ejected quark, in SIDIS by convention denoted by the light-cone $\nminus$ direction. The \TMD correlator obtained from the squared amplitude thus has parallel Wilson lines attached to each of the quark field operators at $0$ and $b$, extending out to infinity along a direction $v \approx \nminus$, see Fig. \ref{fig-link-staple}. Due to the fact that the gauge link is only an effective representation of final state interactions within a framework of suitable approximations, there is a certain degree of freedom with respect to its geometry, in particular with regard to the choice of its direction $v$. At tree level, the most convenient choice is an exactly light-like gauge link, $v = \nminus$. However, going beyond tree-level, it has been found that the light-like link introduces so-called rapidity divergences that are hard to remove, see Ref. \cite{Collins:2007ph} for a review. One way of regulating these divergences is to use a gauge link slightly off the light cone \cite{Collins:1981uw}, see Refs. \cite{Ji:2004wu,Aybat:2011zv,CollinsBook2011} for the application to SIDIS. In Ref. \cite{Ji:2004wu}, the direction $v$ is chosen time-like. More recent work in Refs. \cite{Aybat:2011zv,CollinsBook2011} is based on space-like Wilson lines, motivated by the insight that \TMDs with this choice of link directions feature a ``modified universality'', i.e., they are predicted to be numerically equal for both SIDIS and DY \cite{Collins:2004nx} up to the expected sign changes of T-odd \TMDs. The space-like choice of Wilson lines also opens up the possibility of implementing the gauge link directly in lattice QCD.

In Fig. \ref{fig-link-staple}, the two parallel Wilson lines are connected at the far end by another straight Wilson line. The complete gauge link thus has a staple-like shape. Bridging the transverse gap is necessary to render the operator gauge invariant and proves to be essential if the light-cone gauge $\nminus \cdot A = 0$ is used \cite{Belitsky:2002sm,Brodsky:2002cx}. In a covariant gauge, the connecting link at infinity can be omitted; this has been exploited in Refs.~\cite{Ji:2004wu,Aybat:2011zv}. Lattice calculations are typically performed without any gauge fixing. We therefore prefer the notation with an explicitly gauge invariant operator. Moreover, in our study we take the limit of an infinite ``staple extent'' $\eta$ explicitly. The gauge link employed in this work thus reads
\begin{align}
	\WlineC{\mathcal{C}_\bvec^{(\eta v)}} = \Wline{0, \eta v, \eta v + b, b}\ ,
	\label{eq-staplelink}
\end{align}
where $v$ is space-like. Even at finite $\eta$, this gauge link geometry fulfills the desired symmetry transformation rules, as listed in Eq. (C6) of Ref. \cite{Musch:2010ka} and discussed further below.
Here, we will be mostly concerned with the lowest $x$-moment of \TMDs, corresponding to the case $b^- = b^+ = 0$.
In this case, the connection at the far end is purely transverse.

We choose $v$ space-like and, as in Refs.~\cite{Ji:2004wu,Aybat:2011zv,CollinsBook2011}, we consider \TMDs for the choice $\vprp{v}=0$. 
The Lorentz-invariant quantity characterizing the direction of $v$ is the parameter $\zeta \equiv 2 v \tcdot P/\sqrt{|v^2|}$.
The light-like direction $v = \nminus$ can be approached in the limit $\zeta \rightarrow \infty$.
The parameter $\zeta$ can be understood as an artificial scale or cutoff introduced to regulate rapidity divergences.
Within their work on $e^+e^-$-scattering, Collins and Soper provided evolution equations for the dependence on $\zeta$ applicable for $\zeta \gg \Lambda_{QCD}$ \cite{Collins:1981uk}.
Similar equations have been worked out for all leading-twist and spin-dependent parton distributions \cite{Idilbi:2004vb} based on the formalism of Ref. \cite{Ji:2004wu}.
For the more recent formalism of Refs. \cite{Aybat:2011zv,CollinsBook2011}, evolution equations are presently available for the unpolarized case and the Sivers function \cite{Aybat:2011ge}.
The vectors $P$ and $v$ can be written in terms of rapidities $y_P$ and $y_v$, respectively: 
$P^{\pm} = \mN e^{\pm y_P} / \sqrt{2}$ and $v^+/v^- = - e^{2y_v}$. 
Rewriting $\zeta$ as a dimensionless quantity,
\begin{equation}
	\hat \zeta \equiv \zeta / 2 \mN =  \frac{v \cdot P}{\sqrt{|v^2|} \sqrt{P^2}} = \sinh(y_P - y_v),
	\label{eq-zeta}
\end{equation}
reveals that it is essentially a rapidity difference. 
Notice that the entire system can always be boosted to a frame where $v$ has only spatial components, $v^0 = 0, y_v=0$. This is crucial for the lattice approach.

\subsection{Parametrization of the correlator}
\label{sec-parametr}

The translation of our results obtained in Euclidean space into \TMDs defined and interpreted in the context of light cone coordinates is mediated through a parametrization of the correlator $\tilde \Phi_\unsub$ in terms of manifestly Lorentz-invariant amplitudes.  
For our purposes, it will be important to take the dependence  on the link direction $v$ explicitly into account.
A parametrization of the correlator 
$\Phi_\unsub(\kei,P,S;\infty v,\mu) = \int d^4 \bvec/(2\pi)^4\, e^{i \kei\cdot \bvec}\, \frac{1}{2}\, \bra{P,S}\ \bar{q}(0)\, \Gamma\ \WlineC{\mathcal{C}_\bvec^{(\infty v)}}\, q(\bvec)\ \ket{P,S}$
for link paths that extend to infinity into a direction $v$
has been worked out in Ref. \cite{Goeke:2005hb} and involves 32 independent amplitudes $A_i$ and $B_i$ that depend on the Lorentz-invariant quantities 
$\kei^2$, $\kei \cdot P$, $\kei \tcdot v/v \tcdot P$ and $\hat \zeta$.
Appendix C of Ref. \cite{Boer:2011xd} shows that a parametrization of the corresponding $b$-dependent correlator $\tilde \Phi_\unsub$ is of the same form as the parametrization of $\Phi_\unsub$ if we substitute $\kei \rightarrow - i \mN^2 \bvec$. We thus obtain
\begin{align}
	\frac{1}{2}\widetilde \Phi^{[\Eins]}_{\unsub} & = 
		\mN \tAmp_1 
		- \frac{i \mN^2}{v \tcdot P} \epsilon^{\mu\nu\rho\sigma} P_\mu \bvec_\nu v_\rho S_\sigma \tBmp_5 
		\label{eq-phitildedecomp-scal} \displaybreak[0] \\
	\frac{1}{2}\widetilde \Phi^{[\gamma^5]}_{\unsub} & = 
		\mN^2 (\bvec \tcdot S) \tAmp_5 
		+ \frac{i \mN^2}{P \tcdot v}(v \tcdot S) \tBmp_6 
		\label{eq-phitildedecomp-pscal} \displaybreak[0] \\
	\frac{1}{2}\widetilde \Phi^{[\gamma^\mu]}_{\unsub} & = 
		P^\mu\, \tAmp_2 - i \mN^2 \bvec^\mu\, \tAmp_3
		- i \mN \epsilon^{\mu \nu \alpha \beta} P_\nu \bvec_\alpha S_\beta\, \tAmp_{12} \nonumber \\ &
		+ \frac{\mN^2}{(v \tcdot P)} v^\mu\, \tBmp_1 
		+ \frac{\mN}{v \tcdot P} \epsilon^{\mu \nu \alpha \beta} P_\nu v_\alpha S_\beta\, \tBmp_7  
		- \frac{ i \mN^3}{v \tcdot P} \epsilon^{\mu \nu \alpha \beta} \bvec_\nu v_\alpha S_\beta\, \tBmp_8 \nonumber \\ &
		- \frac{\mN^3}{v \tcdot P} (\bvec \tcdot S) \epsilon^{\mu \nu \alpha \beta} P_\nu \bvec_\alpha v_\beta\, \tBmp_9
		- \frac{i \mN^3}{(v \tcdot P)^2} (v \tcdot S) \epsilon^{\mu \nu \alpha \beta} P_\nu \bvec_\alpha v_\beta \tBmp_{10}
		\label{eq-phitildedecomp-vec} \displaybreak[0] \\ 
	\frac{1}{2}\widetilde \Phi^{[\gamma^\mu \gamma^5]}_{\unsub} & = 
		- \mN S^\mu \tAmp_6 + i \mN (\bvec \tcdot S) P^\mu \tAmp_7 +  \mN^3 (\bvec \tcdot S) b^\mu \tAmp_8 \nonumber \\ &
		+ \frac{i \mN^2}{v \tcdot P} \epsilon^{\mu \nu \rho \sigma} P_\nu \bvec_\rho v_\sigma \tBmp_4
		- \frac{\mN}{v \tcdot P}(v \tcdot S) P^\mu \tBmp_{11}
		+ \frac{i \mN^3}{v \tcdot P}(v \tcdot S) \bvec^\mu \tBmp_{12}  \nonumber \\ &
		+ \frac{i \mN^3}{v \tcdot P}(\bvec \tcdot S) v^\mu \tBmp_{13}
		- \frac{\mN^3}{(v \tcdot P)^2}(v \tcdot S) v^\mu \tBmp_{14} 
		\label{eq-phitildedecomp-pvec} \displaybreak[0] \\
	\frac{1}{2}\widetilde \Phi^{[i \sigma^{\mu\nu} \gamma^5]}_{\unsub} & = 
		i \mN \epsilon^{\mu\nu\rho\sigma} P_\rho \bvec_\sigma \tAmp_4
		+ P^{[\mu} S^{\nu]} \tAmp_9 
		- i \mN^2 \bvec^{[\mu} S^{\nu]} \tAmp_{10} 
		- \mN^2 (\bvec \tcdot S) P^{[\mu} \bvec^{\nu]} \tAmp_{11} \nonumber \\ &
		- \frac{ \mN}{v \tcdot P} \epsilon^{\mu\nu\rho\sigma} P_\rho v_\sigma \tBmp_2
		+ \frac{i\mN^3}{v \tcdot P} \epsilon^{\mu\nu\rho\sigma} \bvec_\rho v_\sigma \tBmp_3
		+ \frac{\mN^2}{v \tcdot P} v^{[\mu} S^{\nu]} \tBmp_{15}
		- \frac{i \mN^2}{v \tcdot P} (\bvec \tcdot S) P^{[\mu} v^{\nu]} \tBmp_{16} \nonumber \\ &
		- \frac{\mN^4}{v \tcdot P} (\bvec \tcdot S) b^{[\mu} v^{\nu]} \tBmp_{17}
		- \frac{i \mN^2}{v \tcdot P} (v \tcdot S) P^{[\mu} b^{\nu]} \tBmp_{18}
		+ \frac{\mN^2}{v \tcdot P} (v \tcdot S) P^{[\mu} v^{\nu]} \tBmp_{19}
		- \frac{i \mN^4}{(v \tcdot P)^2} (v \tcdot S) b^{[\mu} v^{\nu]} \tBmp_{20}
		\label{eq-phitildedecomp-tens}
		 \ ,
\end{align}
where $a^{[\mu}b^{\nu]} \equiv a^\mu b^\nu - a^\nu b^\mu$. The structures above are compatible with the transformation properties of the correlator under the symmetries of QCD. For completeness we list them again in appendix \ref{sec-symtraf}. 

Our previous studies of \TMDs on the lattice \cite{Hagler:2009mb,Musch:2010ka} were carried out with straight gauge links. In that case only the T-even structures involving amplitudes of type $\tAmp_i$ appear in the parametrization. As pointed out already in those references, 
there is not necessarily a one-to-one correspondence between the $A_i$ and $\tAmp_i$ (or the $B_i$ and $\tBmp_i$). 
For example, $\tAmp_8$ contributes to $A_6$, $A_7$ and $A_8$. Note that $l=-\bvec$ in Refs. \cite{Hagler:2009mb,Musch:2010ka}. 

In the above parametrization, factors of $(v \tcdot P)^{-n}$ ensure that the structures are invariant under rescaling of $v$, i.e., $v \rightarrow \alpha v$, for any $\alpha > 0$. 
The above parametrization is therefore suitable for describing the case of the staple links extending to infinity. In that case, only the directional information contained in $v$ should enter. For the lattice calculations, it is however advantageous to start with an equivalent parametrization in which the structures explicitly depend on the staple extent $\eta$ and which is still well-defined for $v \tcdot P = 0$. Such a parametrization can be obtained from the parametrization above by replacing $v \rightarrow \eta v$ and by leaving out the factors $(v \tcdot P)^{-n}$. For example, 
\begin{align}
	\frac{1}{2}\widetilde \Phi^{[\Eins]}_{\unsub}(\bvec,P,S,\eta v,\mu) & = 
		\mN \tilde a_1 
		- i \mN^2 \epsilon^{\mu\nu\rho\sigma} P_\mu \bvec_\nu \eta v_\rho S_\sigma \tilde b_5 
		\label{eq-phitildedecomp-scal-2} \ ,
\end{align}
and analogously for the other Dirac structures. Here we have used lower case amplitudes to distinguish the two parametrizations. The relation to the upper case amplitudes is given by
\begin{align}
	\tAmp_i\left(\bvec^2,\bvec \tcdot P,\frac{v \tcdot \bvec}{v \tcdot P}, \frac{v^2}{(v \tcdot P)^2}, \eta v \tcdot P\right) & = \tilde a_i(\bvec^2,\bvec \tcdot P, \eta v \tcdot b, (\eta v)^2, \eta v \tcdot P)    \, , \nonumber \\
	 \tBmp_i\left(\bvec^2,\bvec \tcdot P,\frac{v \tcdot \bvec}{v \tcdot P}, \frac{v^2}{(v \tcdot P)^2}, \eta v \tcdot P\right) & = (\eta v \tcdot P)^n\ \tilde b_i(\bvec^2,\bvec \tcdot P, \eta v \tcdot b, (\eta v)^2, \eta v \tcdot P)   \, ,
	\label{eq-ab}
\end{align}
where $n$ is the power with which $v \tcdot P$ appears in the denominator in front of the corresponding amplitude $\tBmp_i$ in the parametrization. 
Notice that the $\tilde a_i$ and $\tilde b_i$ are functions of all the Lorentz-invariant products of $\bvec$, $P$ and $\eta v$.
For the upper case amplitudes, however, we choose to represent the dependence on these invariants in the third and fourth argument by $\eta$-independent expressions, in order to facilitate taking the limit $\eta \rightarrow \pm \infty$. 
The dependence on the Collins-Soper parameter $\hat\zeta$ is given by the fourth argument, $v^2/(v \tcdot P)^2 = - 1 /(\mN\hat\zeta)^2$, 
while the fifth argument, $\eta v \tcdot P$, characterizes the length of the gauge link and distinguishes between future and past pointing
Wilson lines. 
For the calculation of \TMDs we work in a frame with $b^+=0$ and $\vprp{v}=\vprp{P}=0$. This leads to a relation that can be expressed in Lorentz-invariant form as
\begin{align}
	\frac{v \tcdot b}{v \tcdot P} = b \tcdot P \frac{R(\zetahat^2)}{\mN^2}\, ,
	\label{eq-lindeprel}
\end{align}
where 
\begin{align}
	R(\zetahat^2) \equiv 1- \sqrt{1 + \hat \zeta^{-2}} = \frac{\mN^2}{v \tcdot P} \frac{v^+}{P^+} \, .
\end{align}
The relation Eq. \eqref{eq-lindeprel} shows that the third argument of the $\tAmp_i$ and $\tBmp_i$ is not independent of the others in the context of \TMDs.
Moreover, in our lattice calculations, we have to choose the link directions $b$ and $v$ such that Eq. \eqref{eq-lindeprel} is  fulfilled. 
As a side remark, the parameter corresponding to Eq. \eqref{eq-lindeprel} in momentum space is $v \tcdot k / v \tcdot P \approx x$, i.e., the amplitudes $A_i$ and $B_i$ acquire an explicit $x$-dependence, which has already been pointed out in Refs. \cite{MuschThesis2,Accardi:2009au}.

For the $\Gamma$-structures at leading twist, the correlator can be written in the form 
\begin{align}
	\frac{1}{2 P^+}\widetilde \Phi^{[\gamma^+]}_{\unsub} & = 	
		\tAmp_{2B} 
		+ i \mN \epsilon_{ij} \vect{\bvec}_i \vect{S}_j\, \tAmp_{12B} 
		\displaybreak[0] \\
	\frac{1}{2 P^+}\widetilde \Phi^{[\gamma^+ \gamma^5]}_{\unsub} & = 
		- \Lambda\, \tAmp_{6B} 
		+ i \left\{ (\bvec \tcdot P) \Lambda - \mN (\vprp{\bvec} \tcdot \vprp{S}) \right\}\, \tAmp_{7B}
		\displaybreak[0] \\
	\frac{1}{2 P^+}\widetilde \Phi^{[i \sigma^{i+} \gamma^5]}_{\unsub} & = 
		i \mN \epsilon_{ij} \vect{\bvec}_j \, \tAmp_{4B} 
		- \vect{S}_i\, \tAmp_{9B} 
		- i\mN \Lambda \vect{\bvec}_i\, \tAmp_{10B}
		+ \mN \left\{ (\bvec \tcdot P) \Lambda - \mN (\vprp{\bvec} \tcdot \vprp{S}) \right\} \vect{b}_i\, \tAmp_{11B}
\end{align}
where the indices $i,j$ correspond to transverse directions, $i,j\in \{1,2\}$
(cf.~appendix~\ref{sec-conv} for further details on notation), and where we
have introduced the following abbreviations for combinations of amplitudes:
\begin{align}
	\tAmp_{2B}  & \ \equiv\  \tAmp_2 + R(\zetahat^2) \tBmp_1 \nonumber \displaybreak[0] \\
	\tAmp_{4B}  & \ \equiv\  \tAmp_4 - R(\zetahat^2) \tBmp_3 \nonumber \displaybreak[0] \\
	\tAmp_{6B}  & \ \equiv\  \tAmp_6 + \left(1-R(\zetahat^2) \right) \left\{ \tBmp_{11} + R(\zetahat^2) \tBmp_{14} \right\} \nonumber \displaybreak[0] \\
	\tAmp_{7B}  & \ \equiv\  \tAmp_7 + R(\zetahat^2) \tBmp_{13} \nonumber \displaybreak[0] \\
	\tAmp_{9B}  & \ \equiv\  \tAmp_9 + R(\zetahat^2) \tBmp_{15} \nonumber \displaybreak[0] \\
	\tAmp_{10B}  & \ \equiv\  \tAmp_{10} - \left(1-R(\zetahat^2) \right) \left\{ \tBmp_{18} - R(\zetahat^2) \tBmp_{20} \right\} \nonumber \displaybreak[0] \\
	\tAmp_{11B}  & \ \equiv\  \tAmp_{11} - R(\zetahat^2) \tBmp_{17} \nonumber \displaybreak[0] \\
	\tAmp_{12B}  & \ \equiv\  \tAmp_{12} - R(\zetahat^2) \tBmp_8 \displaybreak[0] 
	\label{eq-ABamps}
\end{align}
For later convenience we also define
\begin{align}
	\tAmp_{9Bm}  & \ \equiv\  \tAmp_{9B} -  \frac{1}{2} m_N^2 \bvec^2 \widetilde{A}_{11B} \ .
	\label{eq-A9m}
\end{align}
Performing the Fourier transformation and comparing with the decomposition Eqs. \eqref{eq-phigammaplus}-\eqref{eq-phisigmaplusi}, we can express the \TMDs in terms of Fourier-transforms of the above amplitudes.
Using the combined amplitudes $\tAmp_{iB}$,
the results are of the same form as in the straight-link case of Ref. \cite{Musch:2010ka}, 
\begin{align}
	 f_1(x,\vprp{\kei}^2;\zetahat,\ldots,\eta v \tcdot P) & = 2 \fourint\ \widetilde{A}_{2B} \,,\nonumber \displaybreak[0] \\
	 g_1(x,\vprp{\kei}^2;\zetahat,\ldots,\eta v \tcdot P) & = - 2 \fourint\ \widetilde{A}_{6B} + 2 \partial_x \fourint\ \widetilde{A}_{7B} \nonumber \,, \displaybreak[0] \\
	 g_{1T}(x,\vprp{\kei}^2;\zetahat,\ldots,\eta v \tcdot P) & = 4 m_N^2 \partial_{\vprp{\kei}^2} \fourint\ \widetilde{A}_{7B} \nonumber \,, \displaybreak[0] \\
	 h_1(x,\vprp{\kei}^2;\zetahat,\ldots,\eta v \tcdot P) & = - 2 \fourint\  \widetilde{A}_{9Bm} \nonumber \,, \displaybreak[0] \\
	 h_{1L}^\perp(x,\vprp{\kei}^2;\zetahat,\ldots,\eta v \tcdot P) & = 4 m_N^2 \partial_{\vprp{\kei}^2} \left( \fourint \widetilde{A}_{10B} 
	                          + \partial_x \fourint\ \widetilde{A}_{11B} \right) \nonumber \,, \displaybreak[0] \\
	 h_{1T}^\perp (x,\vprp{\kei}^2;\zetahat,\ldots,\eta v \tcdot P) & = 8 m_N^{4} \left(\partial_{\vprp{\kei}^2}\right)^2 \fourint\ \widetilde{A}_{11B}\ ,
	 \label{eq-tmdsfromampseven}
\end{align} 
except that the abbreviation $\fourint$ is now applied to the $v$-dependent amplitudes and includes the soft factor:
\begin{align}
	\fourint \tAmp_i \equiv & \int \frac{d^2 \vprp{\bvec}}{(2\pi)^2}\,e^{-i\vprp{\bvec}\tcdot\vprp{\kei}}\, \frac{1}{\widetilde{\mathcal{S}}(\bvec^2;\ldots)} \int \frac{d(\bvec \tcdot P)}{(2\pi)}\,e^{ix(\bvec \tcdot P)} 
	\tAmp_i(-\vprp{\bvec}^2,\bvec \tcdot P,(\bvec \tcdot P) R(\zetahat^2)/\mN^2,-1/(\mN\zetahat)^2,\eta v \tcdot P) \nonumber \\
	= & \int_0^\infty \frac{d(-\bvec^2)}{2(2\pi)}\ \frac{J_0(\sqrt{-\bvec^2}\, |\vprp{\kei}|)}{\widetilde{\mathcal{S}}(\bvec^2;\ldots)}\ \int \frac{d(\bvec \tcdot P)}{(2\pi)}\,e^{ix(\bvec \tcdot P)}\ 
	\tAmp_i(\bvec^2,\bvec \tcdot P,(\bvec \tcdot P) R(\zetahat^2)/\mN^2,-1/(\mN\zetahat)^2,\eta v \tcdot P) \, 
	\label{eq-fourint}
\end{align}
Also, there are two further \TMDs that are not present in the straight-link case, the T-odd distributions
\begin{align}
	 f_{1T}^\perp(x,\vprp{\kei}^2;\zetahat,\ldots,\eta v \tcdot P) & =  4 m_N^2 \partial_{\vprp{\kei}^2} \fourint\ \widetilde{A}_{12B} \,,\nonumber \displaybreak[0] \\
	 h_1^\perp(x,\vprp{\kei}^2;\zetahat,\ldots,\eta v \tcdot P) & =  - 4 m_N^2 \partial_{\vprp{\kei}^2} \fourint\ \widetilde{A}_{4B}   \, .
	 \label{eq-tmdsfromampsodd}
\end{align} 
Again the dots ``$\ldots$'' indicate further parameters that specify the geometry of the soft factor. 
The T-even distributions $f_1$, $g_1$, $h_1$, $g_{1T}$, $h_{1L}^\perp$ and $h_{1T}^\perp$ fulfill
\begin{align}
	f^{\text{T-even}}(x,\vprp{\kei}^2;\zetahat,\ldots,\eta v \tcdot P) = f^{\text{T-even}}(x,\vprp{\kei}^2;\zetahat,\ldots,-\eta v \tcdot P)
\end{align}
while the T-odd distributions, i.e., at leading twist the Sivers function $f_{1T}^\perp$ and the Boer-Mulders function $h_{1}^\perp$, fulfill
\begin{align}
	f^{\text{T-odd}}(x,\vprp{\kei}^2;\zetahat,\ldots,\eta v \tcdot P) & = - f^{\text{T-odd}}(x,\vprp{\kei}^2;\zetahat,\ldots,-\eta v \tcdot P)
\end{align}
As a result, T-odd distributions must vanish for $\eta = 0$, which corresponds to straight gauge links.
\TMDs for SIDIS and DY are obtained for $\eta v \tcdot P \rightarrow \infty$ and $\eta v \tcdot P \rightarrow -\infty$, respectively.
In the following, we choose $v \tcdot P \geq 0$, such that the SIDIS and DY limits for space-like $v$ can also be written as $\eta |v| \rightarrow \infty$ and $\eta |v| \rightarrow -\infty$, respectively.
Equations \eqref{eq-tmdsfromampseven} and \eqref{eq-tmdsfromampsodd} show that certain $x$-integrated \TMDs in Fourier space directly correspond to the amplitudes $\tAmp_{iB}$ evaluated at $\bvec \tcdot P = 0$ :
\begin{align}
	\tilde f_1^{[1](0)}(\vprp{\bvec}^2;\zetahat,\ldots,\eta v \tcdot P) & = 2\, \widetilde{A}_{2B}(-\vprp{\bvec}^2,0,0,-1/(\mN\zetahat)^2,\eta v\tcdot P)/\widetilde{\mathcal{S}}(\bvec^2;\ldots) \,,\nonumber \displaybreak[0] \\
	\tilde g_1^{[1](0)}(\vprp{\bvec}^2;\zetahat,\ldots,\eta v \tcdot P) & = - 2\, \widetilde{A}_{6B}(-\vprp{\bvec}^2,0,0,-1/(\mN\zetahat)^2,\eta v\tcdot P)/\widetilde{\mathcal{S}}(\bvec^2;\ldots)  \nonumber \,, \displaybreak[0] \\
	\tilde g_{1T}^{[1](1)}(\vprp{\bvec}^2;\zetahat,\ldots,\eta v \tcdot P) & = -2\,  \widetilde{A}_{7B}(-\vprp{\bvec}^2,0,0,-1/(\mN\zetahat)^2,\eta v\tcdot P)/\widetilde{\mathcal{S}}(\bvec^2;\ldots) \nonumber \,, \displaybreak[0] \\
	\tilde h_1^{[1](0)}(\vprp{\bvec}^2;\zetahat,\ldots,\eta v \tcdot P) & = - 2\, \widetilde{A}_{9Bm}(-\vprp{\bvec}^2,0,0,-1/(\mN\zetahat)^2,\eta v\tcdot P)/\widetilde{\mathcal{S}}(\bvec^2;\ldots)  \nonumber \,, \displaybreak[0] \\
	\tilde h_{1L}^{\perp[1](1)}(\vprp{\bvec}^2;\zetahat,\ldots,\eta v \tcdot P) & = -2\,\widetilde{A}_{10B}(-\vprp{\bvec}^2,0,0,-1/(\mN\zetahat)^2,\eta 
	v\tcdot P)/\widetilde{\mathcal{S}}(\bvec^2;\ldots) \nonumber \,, \displaybreak[0] \\
	\tilde h_{1T}^{\perp[1](2)} (\vprp{\bvec}^2;\zetahat,\ldots,\eta v \tcdot P) & = 4\,\widetilde{A}_{11B}(-\vprp{\bvec}^2,0,0,-1/(\mN\zetahat)^2,\eta v\tcdot P)/\widetilde{\mathcal{S}}(\bvec^2;\ldots)\, , \nonumber \displaybreak[0] \\
	\tilde f_{1T}^{\perp[1](1)}(\vprp{\bvec}^2;\zetahat,\ldots,\eta v \tcdot P) & =-2 \,\widetilde{A}_{12B}(-\vprp{\bvec}^2,0,0,-1/(\mN\zetahat)^2,\eta v\tcdot P)/\widetilde{\mathcal{S}}(\bvec^2;\ldots) \,,\nonumber \displaybreak[0] \\
	\tilde h_1^{\perp[1](1)}(\vprp{\bvec}^2;\zetahat,\ldots,\eta v \tcdot P) & =  2 \, \widetilde{A}_{4B}(-\vprp{\bvec}^2,0,0,-1/(\mN\zetahat)^2,\eta v\tcdot P)/\widetilde{\mathcal{S}}(\bvec^2;\ldots)   \, .	
	\label{eq-xintderfttmds}
\end{align} 
The (derivatives of) Fourier-transformed \TMDs $\tilde f_1^{(0)}$, $\tilde g_1^{(0)}$, $\tilde g_{1T}^{(1)}$, $\tilde h_1^{(0)}$, $\tilde h_{1L}^{\perp(1)}$, $\tilde h_{1T}^{\perp(2)}$, $\tilde f_{1T}^{\perp(1)}$ and $\tilde h_1^{\perp(1)}$ 
are naturally accessible from the Fourier-transformed cross section of, e.g., SIDIS \cite{Boer:2011xd}, and naturally appear in evolution equations, see, e.g. \cite{Idilbi:2004vb}.

\subsection{Generalized shifts from amplitudes}

In section \ref{sec-latquan} we have given an example that ratios of certain $\vprp{\kei}$-moments of \TMDs have interesting physical interpretations.
These ratios, and their counterparts generalized to non-zero $\vprp{\bvec}$, are also advantageous from a theoretical point of view:
Obviously, the soft factor $\widetilde{\mathcal{S}}$ cancels in any ratio formed from the objects in Eq. \eqref{eq-xintderfttmds}, along with any
$\Gamma$-independent multiplicative renormalization factor \cite{MuschThesis2,Hagler:2009mb,Musch:2010ka,Boer:2011xd}.

Eq. \eqref{eq-xintderfttmds} identifies $x$-integrated derivatives of Fourier-transformed \TMDs with simple linear combinations of amplitudes $\tAmp_i$ and $\tBmp_i$ evaluated at the same values of $\vprp{\bvec}^2$, $\bvec \cdot P$, $\zetahat$ and $\eta v \tcdot P$. Forming ratios of these objects thus just amounts to taking ratios of linear combinations of the fundamental correlators $\tilde \Phi^{[\Gamma]}_\unsub$ evaluated at the same point, i.e., with the same values for $\bvec$, $P$ and $\eta v$. For a discussion of the renormalization properties of ratios of the objects in \eqref{eq-xintderfttmds} it is thus sufficient to
understand the renormalization properties of (ratios formed from) the
correlators $\widetilde \Phi^{[\Gamma]}_\unsub = \frac{1}{2} \, \bra{P,S}\ \bar{q}(0)\, \Gamma\ \mathcal{U}\ q(\bvec)\ \ket{P,S} $.

Analytical studies of the operator $\bar{q}(0)\, \Gamma\, \mathcal{U} q(\bvec)$ in the continuum \cite{Dotsenko:1979wb,Craigie:1980qs,Arefeva:1980zd,Aoyama:1981ev,Stefanis:1983ke,Dorn:1986dt}
suggest that for $\bvec^2 \neq 0$ the renormalization factors are multiplicative and $\Gamma$-independent. The basic reason is that the quark field operators are at different locations and undergo wave function renormalization separately. We will assume here that our lattice representation of $\bar{q}(0)\, \Gamma\, \mathcal{U}\, q(\bvec)$ is renormalized multiplicatively independent of $\Gamma$ as long as we keep $\vprp{b}^2$ larger than a few lattice spacings. A more detailed discussion and numerical studies of the renormalization properties of this operator can be found in Ref. \cite{Musch:2010ka}. It remains an interesting task for the future to perform a more thorough treatment of non-local operators on the lattice. Under the assumption of multiplicative renormalization, generalized shifts such as $\langle \vect{k}_y \rangle_{TU}(\vprp{\bvec}^2;\zetahat,\eta v \tcdot P) \equiv \mN \tilde f_{1T}^{\perp[1](1)} / \tilde f_1^{[1](0)}$ can only depend on $\vprp{\bvec}^2$, $\zetahat$ and on the staple extent $\eta v \tcdot P$. All other renormalization and soft factor related dependences cancel out in the ratio.
In this work, we will present numerical results for the following generalized shifts:
\begin{align}
	\langle \vect{k}_y \rangle_{TU}(\vprp{\bvec}^2;\zetahat,\eta v \tcdot P) 
	&\ \equiv\ \mN \frac{\tilde f_{1T}^{\perp[1](1)}(\vprp{\bvec}^2;\zetahat,\ldots,\eta v \tcdot P)}{\tilde f_1^{[1](0)}(\vprp{\bvec}^2;\zetahat,\ldots,\eta v \tcdot P)}  
	= - \mN \frac{\widetilde{A}_{12B}(-\vprp{\bvec}^2,0,0,-1/(\mN\zetahat)^2,\eta v\tcdot P) }{ \widetilde{A}_{2B}(-\vprp{\bvec}^2,0,0,-1/(\mN\zetahat)^2,\eta v\tcdot P) } \nonumber \\
	&\ \xrightarrow{\vprp{\bvec}^2 = 0} 
		\left. \frac{ \int dx \int d^2 \vprp{\kei}\, \vect{\kei}_y  \ \Phi^{[\gamma^+]}(x,\vprp{\kei},P,S;\ldots) }
		{ \int dx \int d^2 \vprp{\kei}  \phantom{\vect{\kei}_y}\ \Phi^{[\gamma^+]}(x,\vprp{\kei},P,S;\ldots) } \right|_{\displaystyle \vprp{S}=(1,0) } 
	\label{eq-genSivShift} \displaybreak[0] \\
	\langle \vect{k}_y \rangle_{UT}(\vprp{\bvec}^2;\zetahat,\eta v \tcdot P)
	&\ \equiv\ \mN \frac{\tilde h_{1}^{\perp[1](1)}(\vprp{\bvec}^2;\zetahat,\ldots,\eta v \tcdot P)}{\tilde f_1^{[1](0)}(\vprp{\bvec}^2;\zetahat,\ldots,\eta v \tcdot P)} 
	= \mN \frac{\widetilde{A}_{4B}(-\vprp{\bvec}^2,0,0,-1/(\mN\zetahat)^2,\eta v\tcdot P) }{ \widetilde{A}_{2B}(-\vprp{\bvec}^2,0,0,-1/(\mN\zetahat)^2,\eta v\tcdot P) } \nonumber \\
	&\ \xrightarrow{\vprp{\bvec}^2 = 0} 
		\left. \frac{ \sum_{\Lambda=\pm 1} \int dx \int d^2 \vprp{\kei}\, \vect{\kei}_y  \ \Phi^{[\gamma^+ + s^j i \sigma^{j+}\gamma^5]}(x,\vprp{\kei},P,S;\ldots) }
		{ \sum_{\Lambda=\pm 1} \int dx \int d^2 \vprp{\kei}  \phantom{\vect{\kei}_y}\ \Phi^{[\gamma^+ + s^j i \sigma^{j+}\gamma^5]}(x,\vprp{\kei},P,S;\ldots) } \right|_{\displaystyle \vprp{s}=(1,0) }
	\label{eq-genBMShift} \displaybreak[0] \\
	\langle \vect{k}_x \rangle_{TL}(\vprp{\bvec}^2;\zetahat,\eta v \tcdot P)
	&\ \equiv\  \mN \frac{\tilde g_{1T}^{[1](1)}(\vprp{\bvec}^2;\zetahat,\ldots,\eta v \tcdot P)}{\tilde f_1^{[1](0)}(\vprp{\bvec}^2;\zetahat,\ldots,\eta v \tcdot P)}  =
	- \mN \frac{\widetilde{A}_{7B}(-\vprp{\bvec}^2,0,0,-1/(\mN\zetahat)^2,\eta v\tcdot P) }{ \widetilde{A}_{2B}(-\vprp{\bvec}^2,0,0,-1/(\mN\zetahat)^2,\eta v\tcdot P) } \nonumber \\
	&\ \xrightarrow{\vprp{\bvec}^2 = 0} 
		\left. \frac{ \int dx \int d^2 \vprp{\kei}\, \vect{\kei}_x  \ \Phi^{[\gamma^+ + \lambda \gamma^{+}\gamma^5]}(x,\vprp{\kei},P,S;\ldots) }
		{ \int dx \int d^2 \vprp{\kei}  \phantom{\vect{\kei}_x}\ \Phi^{[\gamma^+ + \lambda \gamma^{+}\gamma^5]}(x,\vprp{\kei},P,S;\ldots) } \right|_{\displaystyle \vprp{S}=(1,0),\ \lambda=1 } 
	\label{eq-geng1TShift} \, ,\\
	\frac{\tilde h_{1}^{[1](0)}(\vprp{\bvec}^2;\zetahat,\ldots,\eta v \tcdot P)}{\tilde f_1^{[1](0)}(\vprp{\bvec}^2;\zetahat,\ldots,\eta v \tcdot P)}  
	&=	- \frac{\widetilde{A}_{9Bm}(-\vprp{\bvec}^2,0,0,-1/(\mN\zetahat)^2,\eta v\tcdot P) }{ \widetilde{A}_{2B}(-\vprp{\bvec}^2,0,0,-1/(\mN\zetahat)^2,\eta v\tcdot P) } 
	\nonumber \\
  &\ \xrightarrow{\vprp{\bvec}^2 = 0} 
		\left.
		 \frac{ \int dx \int d^2 \vprp{\kei}\, \ \Phi^{[s^j i \sigma^{j+}\gamma^5]}(x,\vprp{\kei},P,S;\ldots) }
		{ \int dx \int d^2 \vprp{\kei} \ \Phi^{[\gamma^+]}(x,\vprp{\kei},P,S;\ldots) } \right|_{\displaystyle \vprp{S}=(1,0),\ \vprp{s}=(1,0) } 
	\label{eq-h1overf1} \, .
\end{align}
\begin{itemize}
\item 
The ``generalized Sivers shift'' $\langle \vect{k}_y \rangle^{\text{Sivers}}=\langle \vect{k}_y \rangle_{TU}$ has already been discussed in section \ref{sec-latquan}.
It is T-odd, i.e., we expect to obtain results of opposite sign in the SIDIS and DY limits $\eta v \tcdot P \rightarrow \infty$ and $\eta v \tcdot P \rightarrow -\infty$, respectively.
The generalized Sivers shift describes a feature of the transverse momentum distribution of (unpolarized) quarks in a transversely polarized proton.
In the formal limit $\vprp{\bvec}^2 = 0$ it measures the dipole moment of that distribution orthogonal to the polarization of the proton.
\item 
The ``generalized Boer-Mulders shift'' $\langle \vect{k}_y \rangle^{\text{BM}}=\langle \vect{k}_y \rangle_{UT}$ is also T-odd and addresses the distribution of transversely polarized quarks in an unpolarized proton.
In the limit $\vprp{\bvec}^2 = 0$, the Boer-Mulders shift describes the dipole moment of that distribution orthogonal to the polarization of the quarks.
Note that we use a sum over proton helicities $\sum_{\Lambda=\pm 1}$ in Eq. \eqref{eq-genBMShift} to represent the unpolarized target nucleon.
\item 
The generalized shift $\langle \vect{k}_x \rangle^{\wg}=\langle \vect{k}_x \rangle_{TL}$ 
attributed to the ``worm gear'' function $g_{1T}$ 
quantifies a dipole deformation of the transverse momentum distribution induced by the correlation of the quark helicity and the transverse proton spin.
Unlike the Sivers and the Boer-Mulders shifts, it is a T-even quantity, i.e., the SIDIS and DY limits $\eta v \tcdot P \rightarrow \pm \infty$ are expected to be the same.
This shift has already been studied in lattice QCD using straight gauge links \cite{MuschThesis2,Hagler:2009mb,Musch:2010ka}. 
We are interested to see by how much this ``process independent'' result obtained at $\eta=0$ 
differs from the results calculated with SIDIS- and DY-type gauge links in the limit $\eta v\tcdot P \rightarrow \pm \infty$.
\item 
The ratio $\tilde h_{1}^{[1](0)}/\tilde f_1^{[1](0)}$ can be identified with a ``generalized tensor charge".
Clearly, it is also a T-even quantity, i.e., no differences are expected
between the SIDIS and DY limits $\eta v \tcdot P \rightarrow \pm \infty$.
We have studied $\tilde h_{1}^{[1](0)}$ already in \cite{MuschThesis2,Hagler:2009mb,Musch:2010ka} on the lattice using straight gauge links. 
As $\tilde h_{1}^{[1](0)}/\tilde f_1^{[1](0)}$ doesn't involve any $\vect{k}$-weighting and is directly related to the well-known 
transversity and unpolarized distribution functions, we expect it to be a particularly clean observable.
It therefore qualifies as a very good candidate for our study of the $\eta|v|$-dependence of T-even observables, in particular
the transition from straight to staple-shaped gauge links.
\end{itemize}

The framework laid out above provides the basis for our numerical lattice
calculations described in the next section. Before proceeding, it is
worth reiterating the logic underlying our approach. Recognizing that
the generic kinematics for which \TMDs are defined are space-like, with
light-like separations representing a special limiting case, we proceed
by considering kinematics off the light cone from the start. We again
emphasize that, whereas we thoroughly examine the behavior of our data
as the kinematics are pushed in the direction of the light cone, statements
about formal properties of the light-cone limit lie beyond the
purview of this investigation. Having parametrized the relevant nonlocal
matrix element in terms of Lorentz-invariant amplitudes,
cf.~section~\ref{sec-parametr}, we choose to perform its evaluation
in a Lorentz frame in which the operator under consideration is defined
at one fixed time. There is no obstacle to this choice in view of the
space-like separations entering the original definition of the matrix
element. In this frame, we cast the computation of the matrix element in
terms of a Euclidean path integral, which we evaluate employing lattice
QCD, as detailed in the next section.

\section{Lattice Calculations}

\label{sec-latticecalc}

\subsection{Simulation setup and parameters}

The methodology we use to calculate the non-local correlators on the lattice has been described in detail in Ref. \cite{Musch:2010ka}, except that we now extend this method to staple-shaped links. 
Again, we employ MILC lattices \cite{Ber01,Aubin:2004wf} that have been previously used by the LHP collaboration for GPD calculations \cite{Hagler:2007xi}; however, compared to our previous work with straight gauge links, we now go to lighter pion masses and make use of the coherent proton and anti-proton sequential propagators of Ref. \cite{Bratt:2010jn} to increase our statistics. 
The new LHPC data set offers forward propagators at four different source locations on each gauge configuration. Moreover, coherent proton and antiproton sequential propagators have been calculated, each one implementing simultaneously four nucleon sink locations per gauge configuration. This way it is possible to conduct eight measurements of a three-point function on each gauge configuration in well separated areas of the lattice, boosting statistics significantly. The source-sink separation has been chosen to be nine lattice units. The simulation parameters are summarized in Table \ref{tab-gaugeconfs}.

\begin{table*}[tbp]
	\centering
	\renewcommand{\arraystretch}{1.1}
	\begin{tabular}{|ll|l|c||c|c|c||c|c|}
	\hline
	$\hat m_{u,d}$ & $\hat m_{s}$ & $\hat L^3 \times \hat T$\rule{0ex}{1.2em} & $10/g^2$ & $a\units{(fm)}$  & $m_\pi^\text{DWF}\units{(MeV)}$ & $m_N^\text{DWF}\units{(GeV)}$ & $\#$conf. & $\#$meas. \\
	\hline
	$0.01$ & $0.05$ & $28^3 \times 64$ & $6.76$ & $0.11967(14)(99)$ & $369.0(09)(35)$ & 1.197(09)(12) & 273 & 2184 \\
	$0.01$ & $0.05$ & $20^3 \times 64$ & $6.76$ & $0.11967(14)(99)$ & $369.0(09)(35)$ & 1.197(09)(12) & 658 & 5264 \\
	$0.02$ & $0.05$ & $20^3 \times 64$ & $6.79$ & $0.11849(14)(99)$ & $518.4(07)(49)$ & 1.348(09)(13) & 486 & 3888 \\
	\hline
	\end{tabular}\par\vspace{1ex}
	\renewcommand{\arraystretch}{1.0}
	\caption{Lattice parameters of the $\nf = 2{+}1$ MILC gauge configurations \cite{Ber01,Aubin:2004wf} used in this work. The lattice spacing $a$ has been obtained from the ``smoothed'' values for $r_1/a$ given in Ref. \cite{Bazavov:2009bb} and the value $r_1= 0.3133(26)\units{fm}$ from the analysis of Ref.~\cite{Davies:2009tsa}. The first error estimates statistical errors in $r_1/a$, the second error originates from the uncertainty about $r_1$ in physical units. We also list the pion and the nucleon masses determined in Ref. \cite{Bratt:2010jn} with the LHPC propagators using domain wall valence fermions. The first error is statistical, the second error comes from the conversion to physical units using $a$ as quoted in the table. Note that the masses quoted here in physical units differ slightly from those listed in Refs. \cite{Hagler:2007xi,Bratt:2010jn}, because these references use a different scheme to fix the lattice spacing. The second to last column lists the number of gauge configurations and the last column shows the resulting number of measurements for the calculation of three-point functions achieved by means of multiple locations for source and sink.
	\label{tab-gaugeconfs}%
	}
\end{table*}

\subsection{Nucleon momenta, choice of link directions, and extraction of
amplitudes}
For all of the ensembles listed in Table~\ref{tab-gaugeconfs}, nucleon
momenta $\vect{P} =0$ and $\vect{P} =2\pi/(a\hat{L} ) \cdot (-1,0,0)$,
implemented via corresponding momentum projections in the sequential
propagators, were available. In addition, sequential propagators were
produced corresponding to the nucleon momenta
$\vect{P} =2\pi/(a\hat{L} ) \cdot (-2,0,0)$ and
$\vect{P} =2\pi/(a\hat{L} ) \cdot (1,-1,0)$ for the $\hat{m}_{u,d} =0.02$
ensemble only. We extracted the matrix element
$\widetilde \Phi_\unsub^{[\Gamma]}(\bvec,P,S;\mathcal{C}_\bvec) \equiv
\frac{1}{2}\, \bra{P,S}\ \bar{q}(0)\,
\Gamma\ \WlineC{\mathcal{C}_\bvec}\ q(\bvec)\ \ket{P,S} $ from plateaux
in standard three-point function to two-point function ratios, for a
complete basis of $\Gamma $ structures and nucleon states polarized in the
3-direction. The nucleon momenta $\vect{P} $, quark separations $\vect{b} $
and corresponding staple-shaped gauge link paths $\mathcal{C}_\bvec $ used
on the lattice in the present investigation are listed in
Table~\ref{stapletable}. The link path $\mathcal{C}_\bvec $ is characterized
by the quark separation vector $\vect{b} $ and the staple vector
$\eta \vect{v} $, cf.~Fig.~\ref{fig-links}. The range of $\eta $ studied
was always chosen to extend from zero to well beyond the point where a
numerical signal ceases to be discernible. Furthermore, 
it should be noted that in the case of either $\vect{b} $ or $\vect{v} $
extending into a direction in a lattice plane which forms an angle of
$\pi/4$ with the lattice axes spanning the plane, there are two optimal
approximations of the corresponding continuum path by a lattice link path;
e.g., if one denotes the lattice link vector in $i$-direction as
$\vect{e}_{i} $, then $\vect{b} = 2 (\vect{e}_{1} +\vect{e}_{2} ) $ is
equally well approximated by the sequence of links
$(\vect{e}_{1},\vect{e}_{2},\vect{e}_{1},\vect{e}_{2})$ as by the sequence
$(\vect{e}_{2},\vect{e}_{1},\vect{e}_{2},\vect{e}_{1})$. As far as
$\vect{b} $ is concerned, in such a situation, our calculations
always included both optimal link paths. However, in the case of
$\vect{v} $, in these situations, only one of the two link paths was
included. To be specific, in the instances of
$\eta \vect{v} =\pm n^{\prime } (\vect{e}_{1} \pm \vect{e}_{i} )$ quoted
in Table~\ref{stapletable}, the link path always departs from the quark
locations in $i$-direction, not $1$-direction. This is a shortcoming
of the discretization which breaks the manifest T-transformation
properties present for the continuum staple; presumably it is
responsible for the problematic mixing of T-even and T-odd amplitudes
which we observe in our analysis in the case of staple directions
off the lattice axes. While we expect a symmetry-improved calculation
including both optimal link paths to avoid this issue, with the presently
available data, we find that we need to impose explicitly T-odd/T-even symmetry
in the system of equations from which we extract the amplitudes whenever
$\vect{v} $ does not coincide with a lattice axis.

\begin{table}
\begin{tabular}{|c|c|r|c|}
\hline
$\vect{b} /a $ & $\eta \vect{v} /a $ & $\vect{P} \cdot a\hat{L} /(2\pi )$ &
Notes \\
\hline\hline
$n\cdot (0,0,1), n=-7,\ldots,7$ & $\pm n^{\prime } \cdot (1,0,0)$ &
$(0,0,0)$ & \\
\cline{3-4}
& & $(-1,0,0)$ & \\
\cline{3-4}
& & $(-2,0,0)$ & $\hat{m}_{u,d} =0.02$ ensemble only \\
\cline{2-4}
& $\pm n^{\prime } \cdot (1,1,0)$ & $(-1,0,0)$ & \\
\cline{3-4}
& & $(-2,0,0)$ & $\hat{m}_{u,d} =0.02$ ensemble only \\
\cline{2-4}
& $\pm n^{\prime } \cdot (1,0,0)$ & $(1,-1,0)$ & $\hat{m}_{u,d} =0.02$
ensemble only \\
\cline{2-4}
& $\pm n^{\prime } \cdot (1,-1,0)$ &
$(1,-1,0)$ & $\hat{m}_{u,d} =0.02$ ensemble only \\
\cline{1-4}
$n\cdot (0,1,0), n=-7,\ldots,7$ & $\pm n^{\prime } \cdot (1,0,0)$ &
$(0,0,0)$ & \\
\cline{3-4}
& & $(-1,0,0)$ & \\
\cline{3-4}
& & $(-2,0,0)$ & $\hat{m}_{u,d} =0.02$ ensemble only \\
\cline{2-4}
& $\pm n^{\prime } \cdot (0,0,1)$ & $(-1,0,0)$ & \\
\cline{3-4}
& & $(-2,0,0)$ & $\hat{m}_{u,d} =0.02$ ensemble only \\
\cline{2-4}
& $\pm n^{\prime } \cdot (1,0,1)$ & $(-1,0,0)$ & \\
\cline{3-4}
& & $(-2,0,0)$ & $\hat{m}_{u,d} =0.02$ ensemble only \\
\cline{1-4}
$n\cdot (0,1,1), n=-2,\ldots,2$ & $\pm n^{\prime } \cdot (1,0,0)$ &
$(0,0,0)$ & \\
\cline{3-4}
& & $(-1,0,0)$ & \\
\cline{3-4}
& & $(-2,0,0)$ & $\hat{m}_{u,d} =0.02$ ensemble only \\
\cline{1-4}
$n\cdot (0,-1,1), n=-2,\ldots,2$ & $\pm n^{\prime } \cdot (1,0,0)$ &
$(0,0,0)$ & \\
\cline{3-4}
& & $(-1,0,0)$ & \\
\cline{3-4}
& & $(-2,0,0)$ & $\hat{m}_{u,d} =0.02$ ensemble only \\
\cline{1-4}
$\pm (0,3,\pm 2)$ & $\pm n^{\prime } \cdot (1,0,0)$ & $(0,0,0)$ & \\
\cline{3-4}
& & $(-1,0,0)$ & \\
\cline{3-4}
& & $(-2,0,0)$ & $\hat{m}_{u,d} =0.02$ ensemble only \\
\cline{1-4}
$\pm (0,4,\pm 2)$ & $\pm n^{\prime } \cdot (1,0,0)$ & $(0,0,0)$ & \\
\cline{3-4}
& & $(-1,0,0)$ & \\
\cline{3-4}
& & $(-2,0,0)$ & $\hat{m}_{u,d} =0.02$ ensemble only \\
\cline{1-4}
$\pm (0,4,\pm 3)$ & $\pm n^{\prime } \cdot (1,0,0)$ & $(0,0,0)$ & \\
\cline{3-4}
& & $(-1,0,0)$ & \\
\cline{3-4}
& & $(-2,0,0)$ & $\hat{m}_{u,d} =0.02$ ensemble only \\
\cline{1-4}
$n\cdot (1,1,0), n=-4,\ldots,4$ &
$\pm n^{\prime } \cdot (1,-1,0)$ &
$(1,-1,0)$ & $\hat{m}_{u,d} =0.02$ ensemble only \\
\hline
\end{tabular}
\caption{Sets of staple-shaped gauge link paths and nucleon momenta
$\vect{P} $ used on the lattice. Gauge link paths are characterized by the
quark separation vector $\vect{b} $ and the staple vector $\eta \vect{v}$,
cf.~Fig.~\ref{fig-links}. The surveyed range of $\eta $, parameterized in
the table by the integer $n^{\prime } $, was always chosen to extend from
zero to well beyond the point where a numerical signal ceases to be
discernible. The maximal magnitude of the Collins-Soper parameter
$\hat{\zeta} $ attained in these sets is $|\hat{\zeta} |=0.78$, for
$\vect{P}\cdot a\hat{L}/(2\pi)=(-2,0,0)$ paired with
$\eta \vect{v}/a =\pm n^{\prime } \cdot (1,0,0)$.}
\label{stapletable}
\end{table}

In practice, the overdetermined system of equations which we solve in order
to relate the matrix elements $\widetilde \Phi_\unsub^{[\Gamma]} $ to
the corresponding amplitudes is set up in terms of the quantities
$\tilde{a}_{i} , \tilde{b}_{i} $, cf.~Eq.~(\ref{eq-ab}) in conjunction
with Eqs.~(\ref{eq-phitildedecomp-scal})-(\ref{eq-phitildedecomp-tens}).
This form is suited to include the case $\hat \zeta=0$, where the sign of the
prefactor in front of $\tilde{b}_{i} $ depends on whether the limit
$\eta v \tcdot P = 0$ is approached from the SIDIS or the DY side.

\subsection{Numerical Results}

\subsubsection{The generalized Sivers shift}
In the following, we concentrate on results for the isovector, $u-d$ quark
combination, because in this case contributions from disconnected diagrams and possible vacuum expectation values cancel out.
The errors shown are statistical only. At the present level of accuracy in this exploratory study, we set aside a quantitative analysis of systematic errors. We use the central values for the lattice spacing $a$ as given in Table \ref{tab-gaugeconfs} to convert to physical units. For $\mN$, we consistently substitute the value of the nucleon mass as determined on the lattice, rather than the physical nucleon mass.

\begin{figure}[btp]
	\centering%
	\subfloat[][]{%
		\label{fig-Sivers_lsqr-1_zetasqrlat4}%
		\includegraphics[scale=0.9]{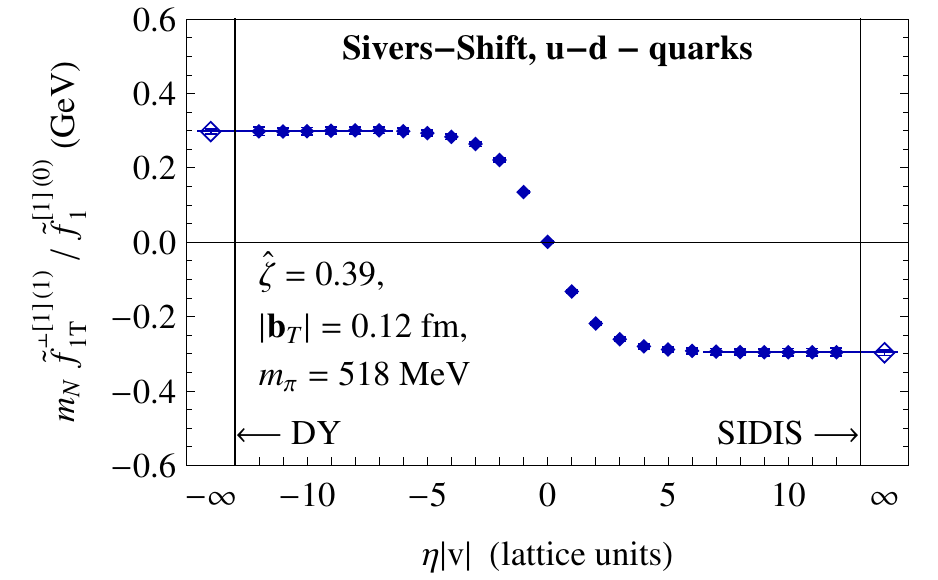}
		}\hfill%
	\subfloat[][]{%
		\label{fig-Sivers_lsqr-4_zetasqrlat4}%\
		\includegraphics[scale=0.9]{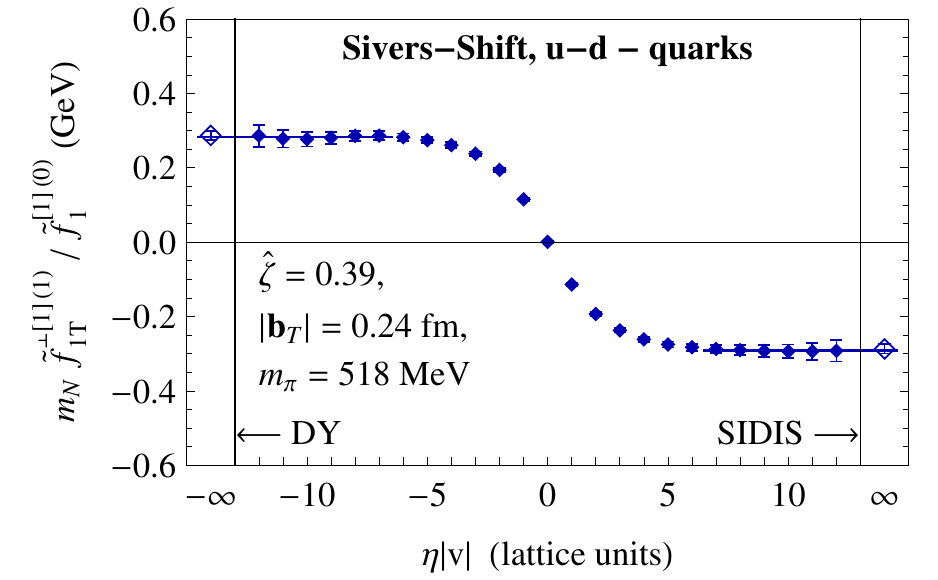}
		}\\%
	\subfloat[][]{%
		\label{fig-Sivers_lsqr-9_zetasqrlat4}%
		\includegraphics[scale=0.9]{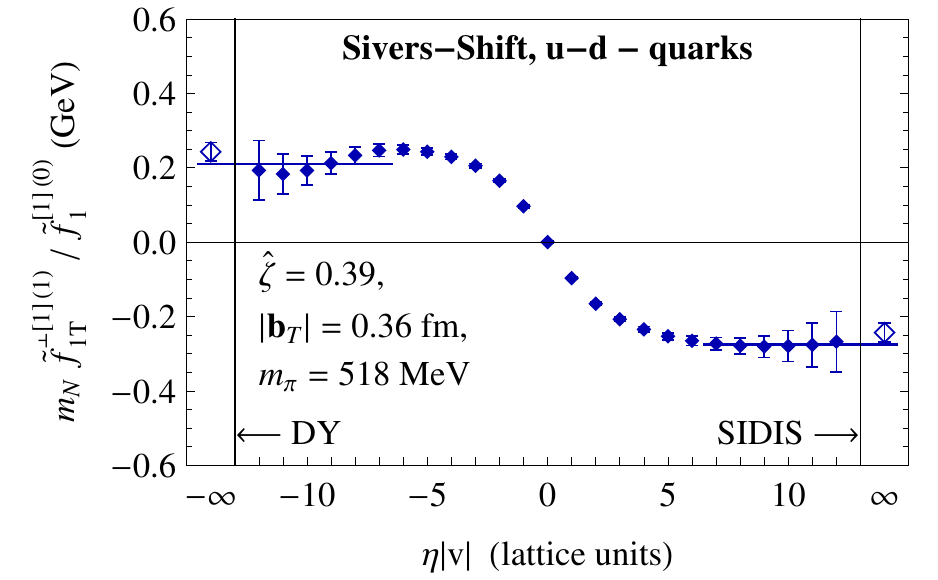}
		}\hfill%
	\subfloat[][]{%
		\label{fig-Sivers_lsqr-16_zetasqrlat4}%\
		\includegraphics[scale=0.9]{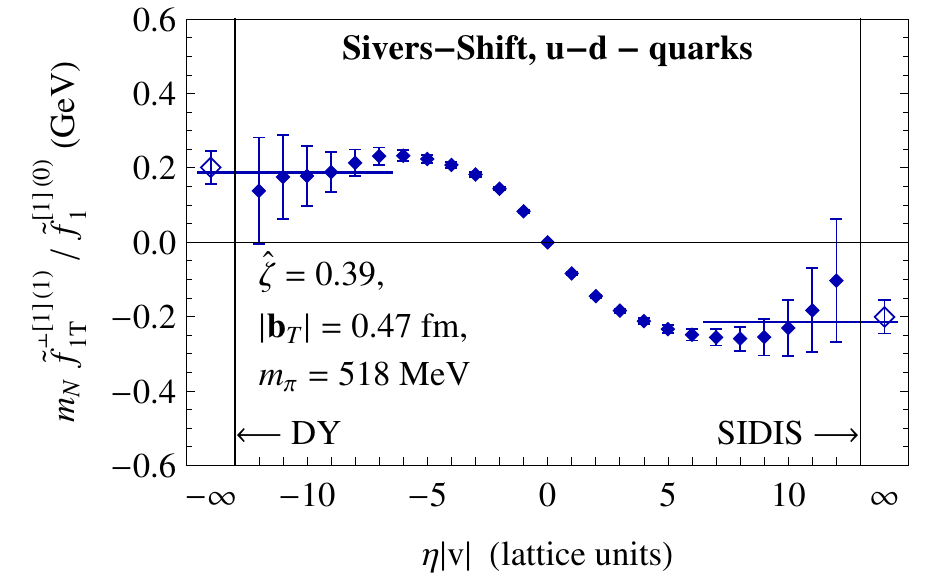}
		}%	
	\caption[SIDIS diagram]{%
		Extraction of the generalized Sivers shift on the lattice with $m_\pi = 518 \units{MeV}$ 
		using a lattice nucleon momentum $|\vect{P}^\lat| = 2\pi /(a\hat{L}) \approx 500 \units{MeV}$ 
		at the corresponding maximal Collins-Soper evolution parameter $\hat \zeta = 0.39$. 
		The continuous horizontal lines are obtained from two independent averages of the data points
		with staple extents in the ranges $\eta |v| = 7a .. 12a$ and $\eta |v| = -12a .. -7a$, respectively. 
		The outer data points shown with empty symbols have been obtained from an anti-symmetrized mean value of these averages,
		i.e., the expected T-odd behavior of the Sivers shift has been put in explicitly. These outer 
		data points are our estimates for the asymptotic values at $\eta |v| \rightarrow \pm \infty$ and thus
		represent the generalized Sivers shifts for SIDIS and DY.
		Error bars show statistical uncertainties only.			
		Figures \subref{fig-Sivers_lsqr-1_zetasqrlat4} and \subref{fig-Sivers_lsqr-4_zetasqrlat4}
		have been obtained with rather small quark field separations $|\vprp{b}|=1a$ and $2a$. 
		Therefore, they might be affected by significant lattice cutoff effects.
		\label{fig-Sivers_etadepend}%
		}
\end{figure}

\begin{figure}[btp]
	\centering%
	\includegraphics[scale=0.9]{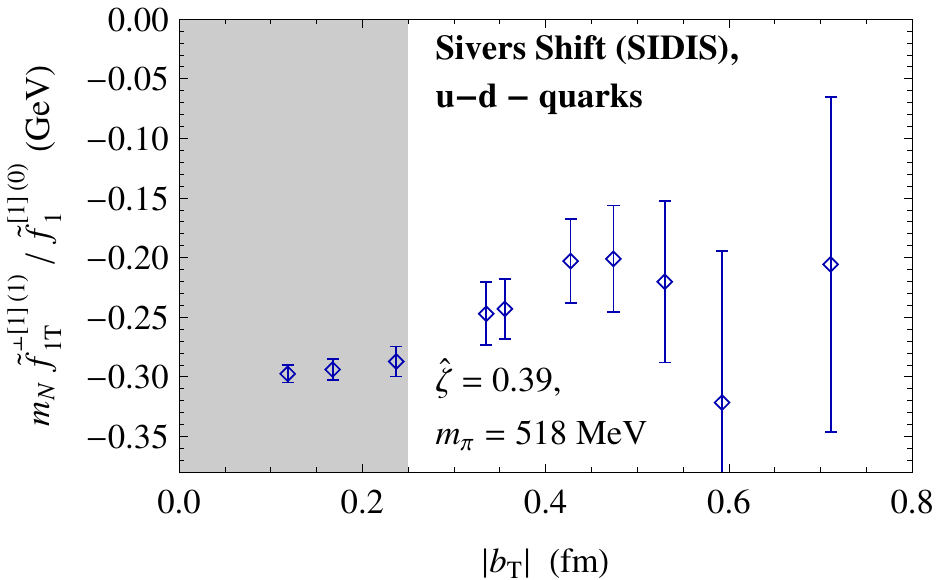}
	\caption{%
		Generalized Sivers shift as a function of the quark separation $|\vprp{b}|$ 
		for the SIDIS case ($\eta |v| = \infty$), extracted
		on the lattice with $m_\pi = 518 \units{MeV}$ for $\hat \zeta = 0.39$.
		The data points lying in the shaded area below $|\vprp{b}| \approx 0.25 \units{fm}$ might be affected by significant lattice cutoff effects.
		Error bars show statistical uncertainties only.
		\label{fig-Sivers_bdepend}%
		}
\end{figure}

\begin{figure}[btp]
	\centering%
	\subfloat[][]{%
		\label{fig-Sivers_lsqr-9_zetasqrlat0}%
		\includegraphics[scale=0.9]{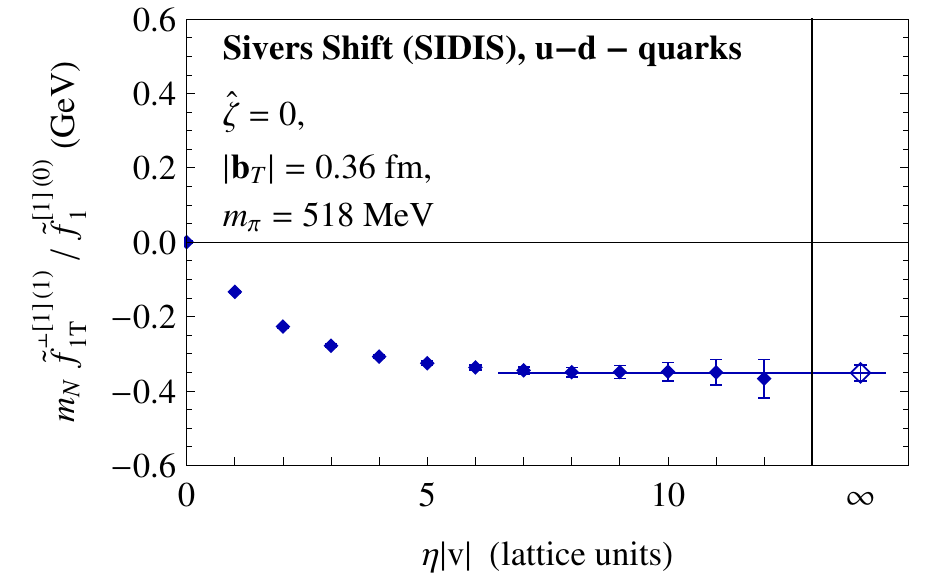}
		}\hfill%
	\subfloat[][]{%
		\label{fig-Sivers_lsqr-9_zetasqrlat16}%\
		\includegraphics[scale=0.9]{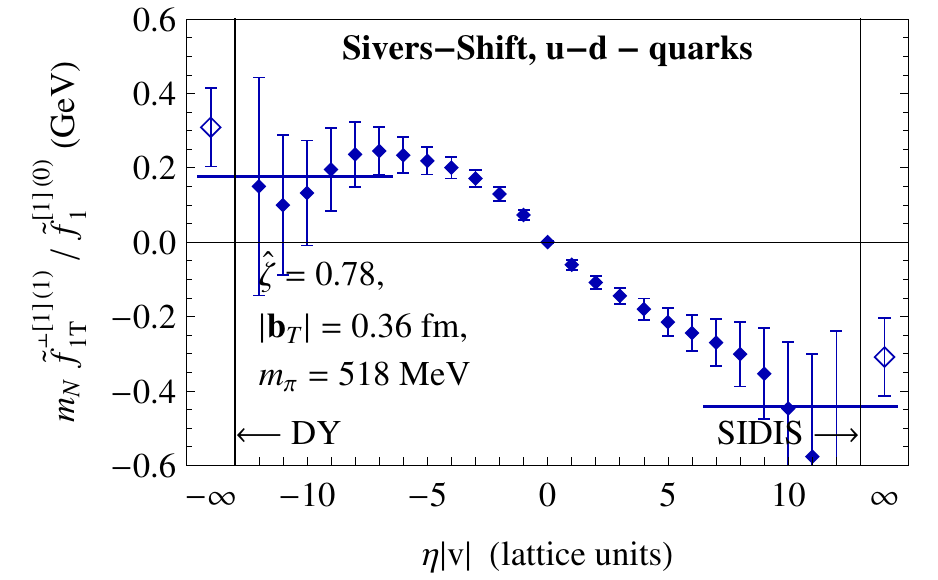}
		}\\%
	\caption[SIDIS diagram]{%
		Generalized Sivers shift on the lattice with $m_\pi = 518 \units{MeV}$ 
		for a quark separation of three lattice spacings, $|\vprp{b}| = 3a = 0.36 \units{fm}$, extracted at $\hat \zeta = 0$
		and at our highest value of the Collins-Soper evolution parameter, $\hat \zeta = 0.78$.
		Figure \subref{fig-Sivers_lsqr-9_zetasqrlat16} has been obtained from nucleons with momentum $|\vect{P}^\lat| = 2 \times 2\pi /(a\hat{L} ) \approx 1 \units{GeV}$ on the lattice. Error bars show statistical uncertainties only.
		\label{fig-Sivers_zetasqrlat_0_16}%
		}
\end{figure}

\begin{figure}[btp]
	\centering%
	\subfloat[][]{%
\label{fig-Sivers_evolution}%\
		\includegraphics[scale=0.9]{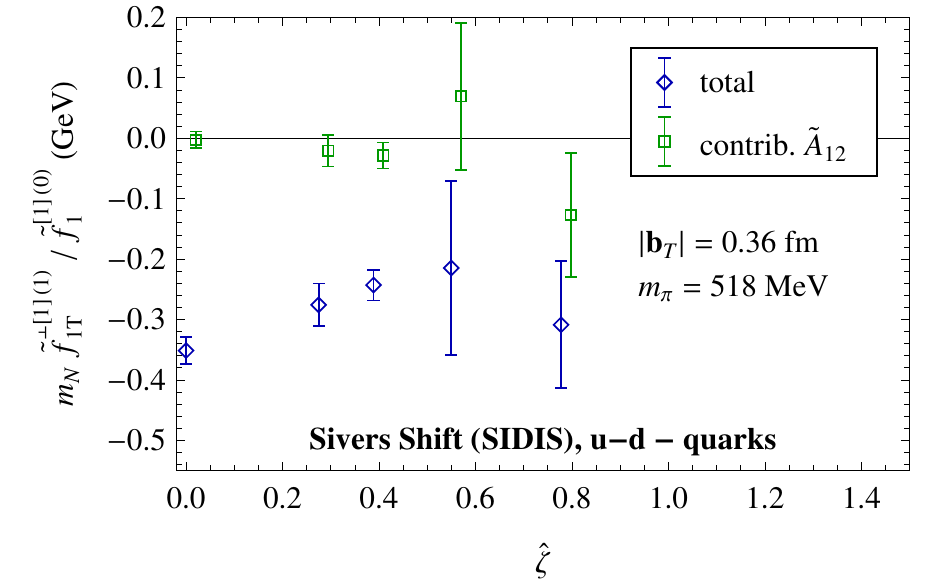}
		}\hfill%
	\subfloat[][]{%
				\label{fig-Sivers_acontrib}%
		\includegraphics[scale=0.9]{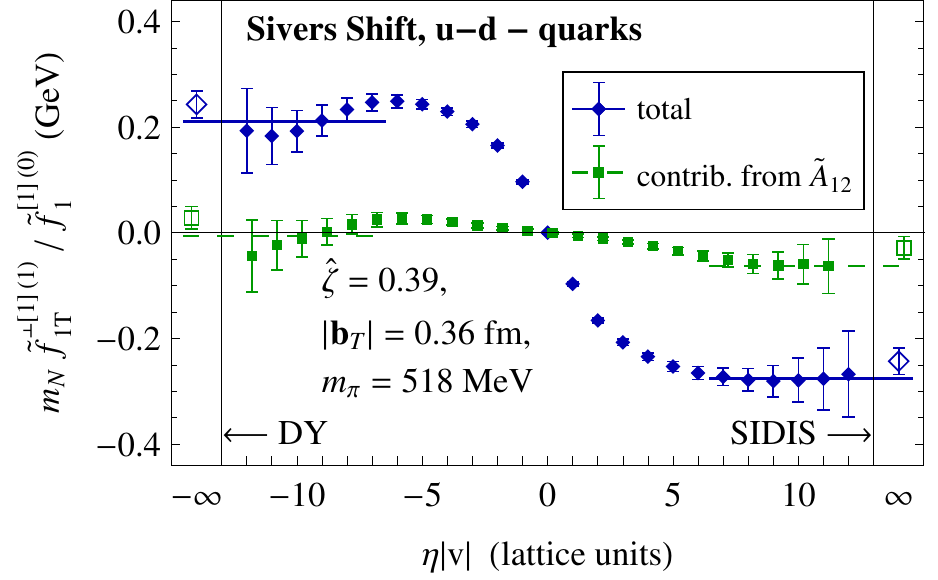}
		}\\%
	\caption[SIDIS diagram]{%
		Generalized Sivers shift on the lattice with $m_\pi = 518 \units{MeV}$ for a quark separation of three lattice spacings, $|\vprp{b}| = 3a =0.36 \units{fm}$.
		In Figure \subref{fig-Sivers_evolution} we show the $\hat \zeta$-dependence of the generalized Sivers shift, depicting both the full result and the result obtained 
		with just $\tAmp_{12}$ in the numerator. The data points correspond to those displayed in the SIDIS limit $\eta |v| \rightarrow \infty$ in plots such as Fig. \subref{fig-Sivers_acontrib}. 
		Figure \subref{fig-Sivers_acontrib} shows the $\eta$-dependence at  $\hat \zeta = 0.39$ for both the full result (diamonds) 
		and the contribution from amplitude $\tAmp_{12}$ in the numerator (squares).
		Asymptotic results corresponding to SIDIS and DY have been extracted as in Fig. \ref{fig-Sivers_etadepend}.
		Error bars show statistical uncertainties only.
		\label{fig-Sivers-aevolution}%
		}
\end{figure}

\begin{figure}[btp]
	\centering%
	\includegraphics[scale=0.9]{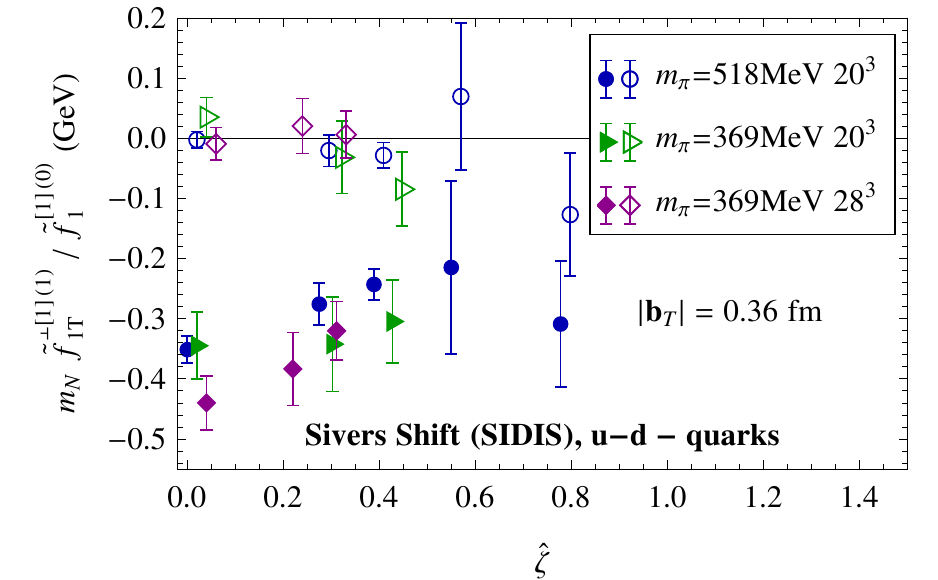}
	\caption{%
		Comparison of the $\hat \zeta$-evolution of the generalized Sivers shift at $|\vprp{b}|=3a=0.36\units{fm}$ 
		for the three different lattices listed in Table \ref{tab-gaugeconfs}.
		Filled symbols correspond to the full SIDIS result. The data points with open symbols have been 
		obtained with only $\tAmp_{12}$ in the numerator. Error bars show statistical uncertainties only.
		\label{fig-Sivers_evolution_combined}%
		}
\end{figure}

Figures \ref{fig-Sivers_etadepend} to \ref{fig-Sivers_evolution_combined} show our results for the generalized Sivers shift, $\langle \vect{k}_y \rangle_{u-d}^{\text{Sivers}}$.
We begin with a discussion of its dependence on the staple orientation, i.e., SIDIS- or DY-like, and the staple extent, $\eta |v|$, 
as displayed in Fig.~\ref{fig-Sivers_etadepend} for a Collins-Soper evolution parameter of
$\hat \zeta = 0.39$ and a pion mass of $m_\pi = 518 \units{MeV}$.
As mentioned before, the T-odd Sivers function must vanish for $\eta |v|=0$, i.e., a straight Wilson line between the quark fields,
but non-vanishing results are allowed (and generally expected) for non-zero staple extents. 
Furthermore, the T-odd observables are anti-symmetric in $\eta |v|$, so we expect the Sivers shift to be of the same size but
opposite in sign for the SIDIS and the DY cases. 
This is exactly what we find in, e.g., Fig.~\ref{fig-Sivers_lsqr-1_zetasqrlat4}, showing the shift for a quark-antiquark distance of 
a single lattice spacing, $|\vprp{b}|=1a$. 
The aforementioned features are realized in form of a curve that is reminiscent of a hyperbolic tangent.
We stress that the observed zero crossing with a change in sign is directly caused by the underlying gauge-invariant operators and their
symmetry properties, and hence represents a consistency check of our calculation rather than any sort of a prediction.

Remarkably, already as $|\eta| |v|$ approaches values of $\sim 6a$, we find that the Sivers shift stabilizes and reaches specific plateau values.
Apart from finite volume effects, in particular wrap-around effects due to the periodic boundary conditions on the lattice, 
we see no reason to expect that once a plateau has been reached, 
the value of the shift would significantly change as $|\eta| |v|\rightarrow\infty$.
To obtain first estimates for staple-shaped Wilson lines that have an infinite extent in $v$-direction,
we therefore choose to average the shifts in the plateau regions $|\eta| |v|=7a \ldots 12a$, as illustrated by the straight lines. 
Clearly, as $|\vprp{b}|$ increases from $0.12\units{fm}$ in Fig.~\ref{fig-Sivers_lsqr-1_zetasqrlat4} to 
$0.47\units{fm}$ in Fig.~\ref{fig-Sivers_lsqr-16_zetasqrlat4}, the signal-to-noise ratio decreases as we approach larger values of $|\eta| |v|$.
For smaller $|\eta| |v|$ 
the statistical uncertainties are much smaller, and the corresponding values tend to dominate the averages when 
the errors are taken into account as weights.
At the same time, however, these statistically dominating data points are more likely to introduce systematic uncertainties related to the (unknown) onset of the ``true" plateau region and the corresponding starting value for the averaging procedure.
Therefore, in order to avoid a too strong bias from the data at smaller $|\eta| |v|$,
we do not use the respective statistical errors as weights in the averaging.
Our final estimates for the Sivers shift are obtained from the mean value of the SIDIS and DY averages and by imposing antisymmetry
in $\eta |v|$. The results are displayed as open diamonds at $\eta |v|=\pm\infty$ in Fig.~\ref{fig-Sivers_etadepend}.
The dependence of these results on $|\vprp{b}|$ is shown in Fig.~\ref{fig-Sivers_bdepend}.
In summary, for $\hat \zeta = 0.39$ and $|\vprp{b}|=0.12\ldots0.47\units{fm}$, we find a sizeable negative Sivers shift for $u-d$ quarks
in the range of $\langle \vect{k}_y \rangle_{u-d}^{\text{Sivers,SIDIS}}=-0.3\ldots-0.15 \units{GeV}$.

Next, we turn to the dependence of our results on the Collins-Soper evolution parameter $\hat\zeta$.
In Fig.~\ref{fig-Sivers_zetasqrlat_0_16}, we consider two ``extreme" cases, namely, a vanishing $\hat\zeta$ as well as the largest
$\hat\zeta=0.78$ that we could access in this study.
While we find rather precise values for the Sivers shift for $\hat\zeta=0$ with a well-defined plateau\footnote{Note that, in the case at hand, $\hat\zeta=0$
corresponds to $\vect{P}=0$, so that one cannot identify a ``forward'' or ``backward'' direction.
Hence, there is only a single branch in $\eta |v|$, the sign of which is a matter of definition, see also the discussion further below in the text.}
for $|\eta| |v|\ge 6a$ in Fig.~\ref{fig-Sivers_lsqr-9_zetasqrlat0}, fluctuations and uncertainties 
quickly increase with $|\eta||v|$ for $\hat\zeta=0.78$ in Fig.~\ref{fig-Sivers_lsqr-9_zetasqrlat16}.
In particular, it is difficult to identify the onset of a plateau on the right hand (SIDIS) side of Fig.~\ref{fig-Sivers_lsqr-9_zetasqrlat16}.
Following the averaging procedure described above, we however find that the estimated values at $|\eta| |v|=\infty$ for the two extreme cases of $\hat\zeta$ 
agree within uncertainties.

Figure \ref{fig-Sivers_evolution} shows the Sivers shift as a function of $\hat\zeta$, for $|\vprp{b}| =0.36 \units{fm}$
and a pion mass of $m_\pi = 518 \units{MeV}$. 
Within the present uncertainties, we observe a statistically significant negative shift; however, it is not possible to 
identify a clear trend of the data points as $\hat\zeta$ increases.
With respect to data points obtained for staple link directions $\vect{v}$ off the lattice axes, 
i.e., $\hat \zeta\approx 0.55$ in Fig.~\ref{fig-Sivers_evolution}, 
we note again that we need to impose the T-odd/T-even (anti-)symmetry already when we solve our system of equations,
in order to avoid problematic mixings of T-even and T-odd amplitudes. 
As already mentioned further above, we
expect this to become unnecessary in the case that lattice symmetry improved operators (see Appendix D of 
Ref.~\cite{Musch:2010ka}) are used.

We find it very interesting to note that the contribution from $\tilde A_{12}$ alone in the numerator of Eq. \eqref{eq-genSivShift} (rather than $\tilde A_{12B}$), illustrated by the open squares,
is essentially compatible with zero within errors for all accessible values of $\hat \zeta$. 
The main contribution to the transverse shift therefore comes from $-R(\hat \zeta^2)\tilde B_{8}=-\eta\,(v\cdot P)\,R(\hat \zeta^2)\tilde b_8$ 
(see Eqs.~(\ref{eq-ab})), i.e., the amplitude $\tilde b_8$. 
Note again that, on the lattice, we employ expressions in terms of the lower-case $\tilde a_i$ and $\tilde b_i$ amplitudes, e.g. $\tilde b_8$, as they are well defined even when $\hat \zeta\rightarrow0$. 
In this limit, $v\cdot P\rightarrow0$, and hence the prefactor behaves as $-(v\cdot P)\,R(\hat \zeta^2)\rightarrow a \mN$. 
The sign of the prefactor of $\tilde b_i$ depends on whether one approaches the limit $v\cdot P\rightarrow0$ from the SIDIS or the DY side.

An example that explicitly shows the relative smallness of $\tilde A_{12}$ is given in Fig.~\ref{fig-Sivers_acontrib}, 
for $\hat \zeta = 0.39$ and $|\vprp{b}| =0.36 \units{fm}$.
While $\tilde A_{12}$ as a function of $\eta |v|$ shows the typical behavior expected for a T-odd amplitude, 
it represents only about $10\%$ of the total contribution for, e.g., $|\eta||v|=6a$.

As one of our central results, we show in Fig.~\ref{fig-Sivers_evolution_combined} 
the Sivers shift as a function of $\hat \zeta$ for all considered ensembles, as before for a fixed $|\vprp{b}| =0.36 \units{fm}$.
Within statistical uncertainties, the data points for the two different pion masses $m_\pi = 369 \units{MeV}$ and $m_\pi = 518 \units{MeV}$,
as well as the spatial lattice volumes $V\approx(2.4 \units{fm})^3$ and $V\approx(3.4 \units{fm})^3$, are overall well compatible.
Apart from the less well determined data point at $\hat\zeta\approx 0.55$, we find
a clearly non-zero negative Sivers shift in the range
$\langle \vect{k}_y \rangle_{u-d}^{\text{Sivers,SIDIS}}=-0.48\ldots-0.2 \units{GeV}$.
Together with the relatively mild $\vprp{b}$-dependence at smaller $\vprp{b}$, cf.~Fig.~\ref{fig-Sivers_bdepend},
this provides strong evidence that the ($x$- and $\vprp{k}$-moment of the)
Sivers function $f_{1T}^{\perp}$ considered here is sizeable and negative
for $u-d$ quarks.
Our preliminary separate data for $u$- and for $d$-quarks (not shown in this work) furthermore indicate that 
$f_{1T}^{\perp,u}<0$ and $f_{1T}^{\perp,d}>0$.
Although our results for the T-odd Sivers effect are still subject to many systematic effects and uncertainties, it is
interesting to note that they are overall well compatible with results from a phenomenological analysis of SIDIS data 
\cite{Anselmino:2005ea,Anselmino:2011gs}, as well as arguments based on the chromodynamic lensing mechanism by Burkardt \cite{Burkardt:2002ks,Burkardt:2003je,Burkardt:2003uw}.
It should also be noted that, in a recent twist-3 analysis of single spin asymmetries from RHIC experiments,
a possible discrepancy has been found with respect to the signs \cite{Kang:2011hk}.

We stress again that fully quantitative predictions for, or a comparison with, 
phenomenological and experimental TMD studies employing QCD factorization
would require lattice data for much larger Collins-Soper parameters,
$\hat\zeta \gg 1$.
With $\hat \zeta^2=- (v \tcdot P)^2/(v^2\mN^2)$, the limit $\hat\zeta\rightarrow\infty$ 
corresponds to the limit of a light-like staple direction $v$,
or an infinite rapidity $y_v\rightarrow -\infty$ in Eq.~(\ref{eq-zeta}). 
For the shifts and ratios defined in Eqs.~(\ref{eq-genSivShift})-(\ref{eq-h1overf1}), 
where the soft factors in the TMD definitions \cite{Ji:2004wu,Aybat:2011zv,CollinsBook2011}
cancel out, large values of $\hat\zeta$ can also be accessed through large nucleon momenta.
Clearly, the limit of an infinite Collins-Soper parameter is in practice not accessible on the lattice, so that
we have to rely on results for a limited range of $\hat \zeta$, as for example in Fig.~\ref{fig-Sivers_evolution_combined}.
From the perturbative prediction for the $\zeta$ dependence, cf., e.g., Ref.~\cite{Aybat:2011zv}, we would expect that 
the ratios of TMDs should become independent of $\hat\zeta$ as $\hat\zeta\rightarrow\infty$.
It would be very interesting to investigate this on the basis of future lattice results for 
larger hadron momenta and with substantially improved statistics.  

\subsubsection{The generalized Boer-Mulders shift}
\begin{figure}[btp]
	\centering%
	\subfloat[][]{%
		\label{fig-BoerMuld_lsqr-9_zetasqrlat4}%
		\includegraphics[scale=0.9]{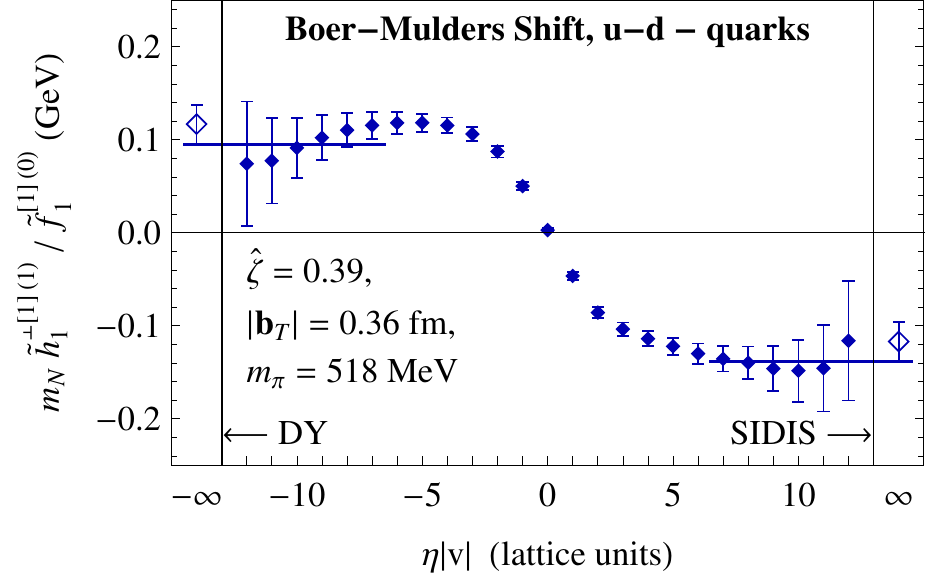}
		}\hfill%
	\subfloat[][]{%
		\label{fig-BoerMuld_bdepend}%\
		\includegraphics[scale=0.9]{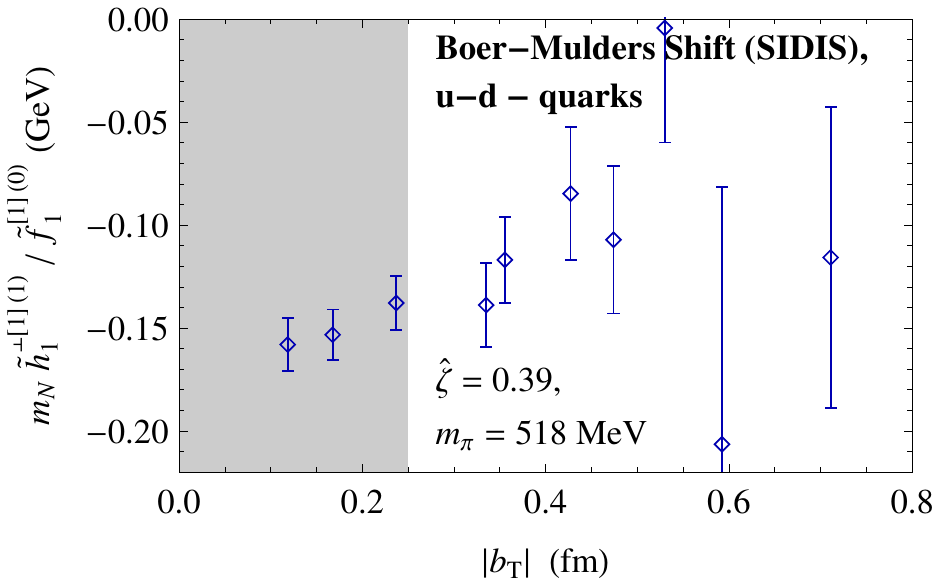}
		}\\%
	\subfloat[][]{%
		\label{fig-BoerMuld_evolution}%
		\includegraphics[scale=0.9]{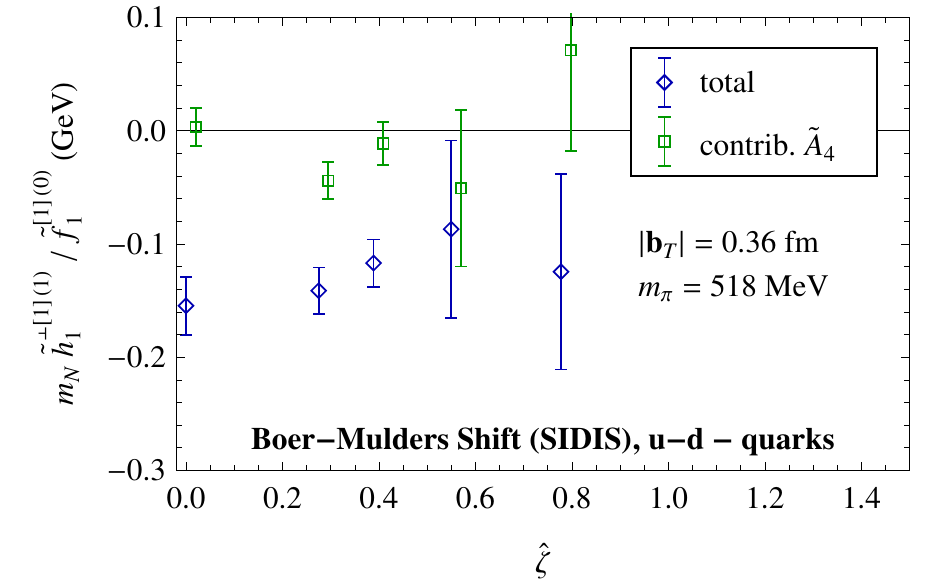}
		}\hfill%
	\subfloat[][]{%
		\label{fig-BoerMuld_evolution_combined}%\
		\includegraphics[scale=0.9]{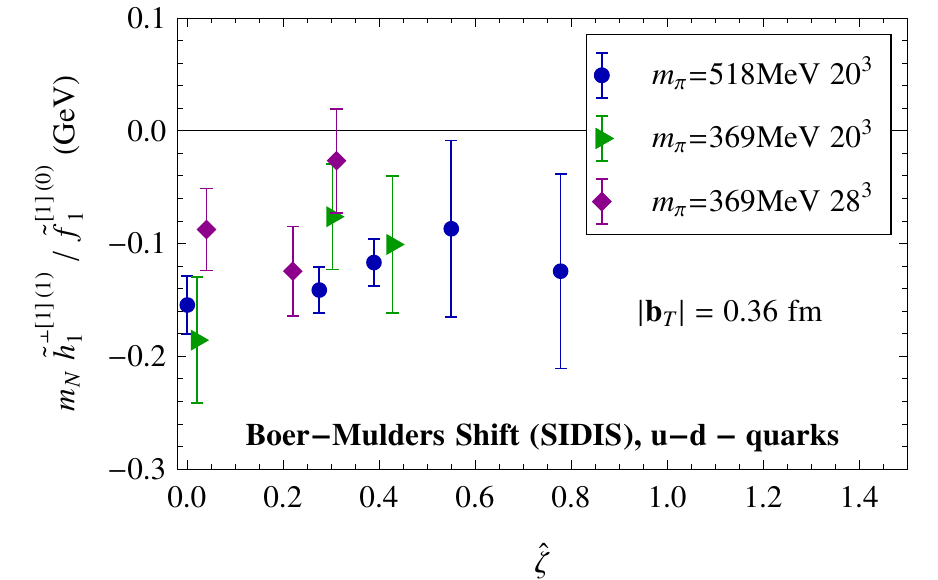}
		}%	
	\caption[SIDIS diagram]{%
		Generalized Boer-Mulders shift.\par
		\subref{fig-BoerMuld_lsqr-9_zetasqrlat4} 
			$\eta|v|$-dependence at $m_\pi = 518 \units{MeV}$ for $\hat \zeta = 0.39$, $|\vprp{b}|=3a=0.36\units{fm}$. \\
		\phantom{\subref{fig-BoerMuld_lsqr-9_zetasqrlat4}} Asymptotic results corresponding to SIDIS and DY have been extracted as in Fig. \ref{fig-Sivers_etadepend}.\par
		\subref{fig-BoerMuld_bdepend} 
			$|\vprp{b}|$-dependence of the SIDIS results at $m_\pi = 518 \units{MeV}$, $\hat \zeta = 0.39$. \par
		\subref{fig-BoerMuld_evolution} 
			$\hat \zeta$-dependence of the SIDIS results at $m_\pi = 518 \units{MeV}$, $|\vprp{b}|=0.36\units{fm}$. \\
		\phantom{\subref{fig-BoerMuld_bdepend}} Empty squares correspond to the ratio with $\tAmp_{4}$ in the numerator only. \par
		\subref{fig-BoerMuld_evolution_combined} 
			Comparison of the $\hat \zeta$-dependence of the SIDIS results obtained from the three different lattice ensembles listed in Table \ref{tab-gaugeconfs}. \par	
		All error bars show statistical uncertainties only.
		\label{fig-BoerMuld}%
		}
\end{figure}

We now turn to the second prominent T-odd TMD, the Boer-Mulders function.  
Our results for the generalized Boer-Mulders shift $\langle \vect{k}_y \rangle_{u-d}^{\text{BM}}$ (Eq.~(\ref{eq-genBMShift}))
are summarized in Figs.~\ref{fig-BoerMuld_lsqr-9_zetasqrlat4} to \ref{fig-BoerMuld_evolution_combined}.
A typical example for the $\eta |v|$-dependence is shown in Fig.~\ref{fig-BoerMuld_lsqr-9_zetasqrlat4}
for a pion mass of $m_\pi = 518 \units{MeV}$, $\hat \zeta = 0.39$, and $|\vprp{b}|=0.36\units{fm}$.
Apart from the magnitude of the shift, the results are very similar to what we have found for the Sivers shift
in Fig.~\ref{fig-Sivers_lsqr-9_zetasqrlat4} above, with indications for plateaus for $|\eta| |v|\ge 6a$.
Figure \ref{fig-BoerMuld_bdepend} illustrates the dependence on $|\vprp{b}|$ for the SIDIS case.
Although the central values indicate some trend towards values smaller in magnitude as $|\vprp{b}|$ increases, 
the somewhat large uncertainties and fluctuations at larger $|\vprp{b}|$ prevent us from drawing any strong conclusions.
In the range of $|\vprp{b}|\approx 0\ldots 0.4 \units{fm}$, we find a clearly non-zero negative Boer-Mulders shift
of $\langle \vect{k}_y \rangle_{u-d}^{\text{BM,SIDIS}}\approx -0.17\ldots-0.1 \units{GeV}$, for $\hat \zeta = 0.39$ and the given pion mass.
The $\hat\zeta$-dependence for $m_\pi = 518 \units{MeV}$ and $|\vprp{b}|=0.36\units{fm}$ is shown in Fig.~\ref{fig-BoerMuld_evolution}.
As for the Sivers shift, it is interesting to note that the contribution from the $\tilde A$ amplitude, 
in this case $\tilde A_4 $, given by the open squares, is mostly
compatible with zero within errors, while the main signal is coming from $-R(\zetahat^2) \tBmp_3=-\eta\,(v\cdot P)\,R(\hat \zeta^2)\tilde b_3$,
cf.~Eqs.~(\ref{eq-ABamps}).

Finally, a comparison of the results and their $\hat\zeta$-dependences for the three different lattice ensembles is provided in Fig.~\ref{fig-BoerMuld_evolution_combined}, for a fixed $|\vprp{b}|=0.36\units{fm}$. 
We find that most of the data points for the two pion masses and the two volumes are well compatible within uncertainties,
with central values of $\langle \vect{k}_y \rangle_{u-d}^{\text{BM,SIDIS}}\approx -0.2\ldots-0.1 \units{GeV}$.
While the central values show little dependence on $\hat\zeta$, the errors have to be significantly reduced before
any extrapolations towards a large Collins-Soper parameter may be attempted.
In summary, for the given ranges of parameters, our results indicate that the Boer-Mulders function is sizeable and negative for $u-d$ quarks.
Our data for the individual $u$- and $d$-quark contributions (not shown) furthermore indicate that $h^{\perp,u}_1<0$ and $h^{\perp,d}_1<0$.
Interestingly, these preliminary results are well compatible with a recent phenomenological study of the Boer-Mulders effect in SIDIS \cite{Barone:2009hw}, as well as an earlier lattice QCD study of tensor generalized parton distributions \cite{Gockeler:2006zu} in combination
with the chromodynamic lensing mechanism \cite{Burkardt:2005hp}.

\subsubsection{$T$-even TMDs: The transversity $h_1$}
In the previous sections, we have discussed the T-odd Sivers and Boer-Mulders distributions, in particular their emergence
in the transition from straight to staple-shaped gauge links, i.e., as $\eta|v|$ changes from zero to large positive or negative values.
A natural question to ask is, what is the influence of final state interactions, which we
mimick on the lattice with the staple-shaped links, and which are essential for the appearance of T-odd distributions, on the T-even TMDs?
More specifically, we would like to see whether and how the T-even distributions, which are generically non-vanishing already for straight
gauge links, change during the transition to finite staple extents.
This is also of considerable interest with respect to the much less involved lattice studies of (T-even) TMDs using straight gauge links 
that we have presented in \cite{Musch:2010ka}.
As we will show, there is only little difference in the transition to staple-shaped links, such that 
our previous results might be of greater phenomenological 
importance than initially expected for the straight ``process-independent" gauge link structures.

\begin{figure}[btp]
	\centering%
	\subfloat[][]{%
		\label{fig-h1Rat_lsqr-1_zetasqrlat4}%
		\includegraphics[scale=0.9]{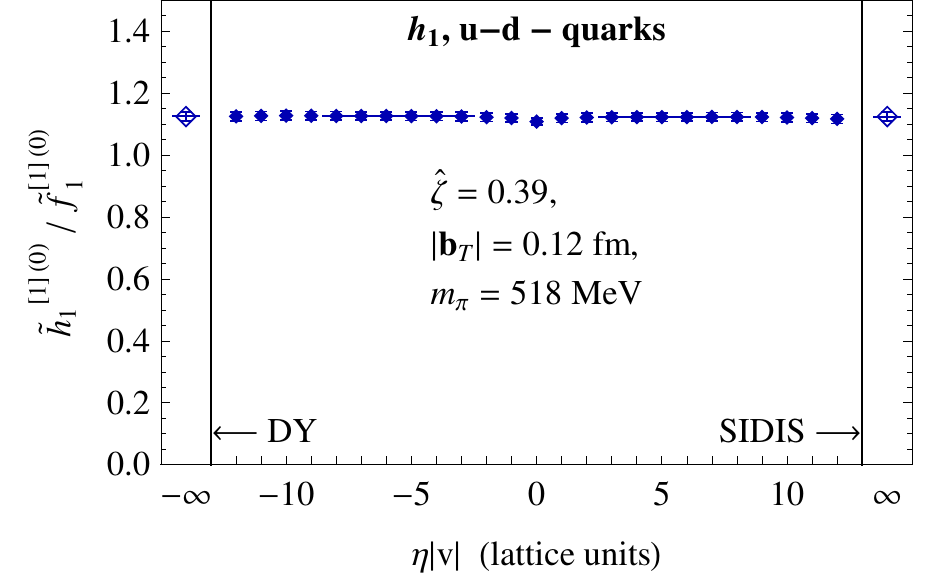}
		}\hfill%
	\subfloat[][]{%
		\label{fig-h1Rat_lsqr-4_zetasqrlat4}%\
		\includegraphics[scale=0.9]{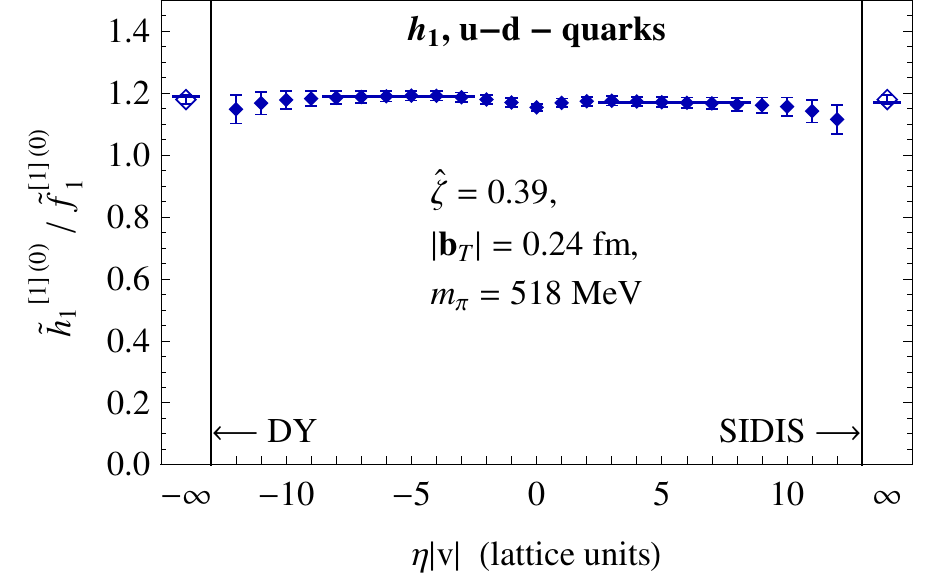}
		}\\%
		\subfloat[][]{%
		\label{fig-h1Rat_lsqr-9_zetasqrlat4}%\
		\includegraphics[scale=0.9]{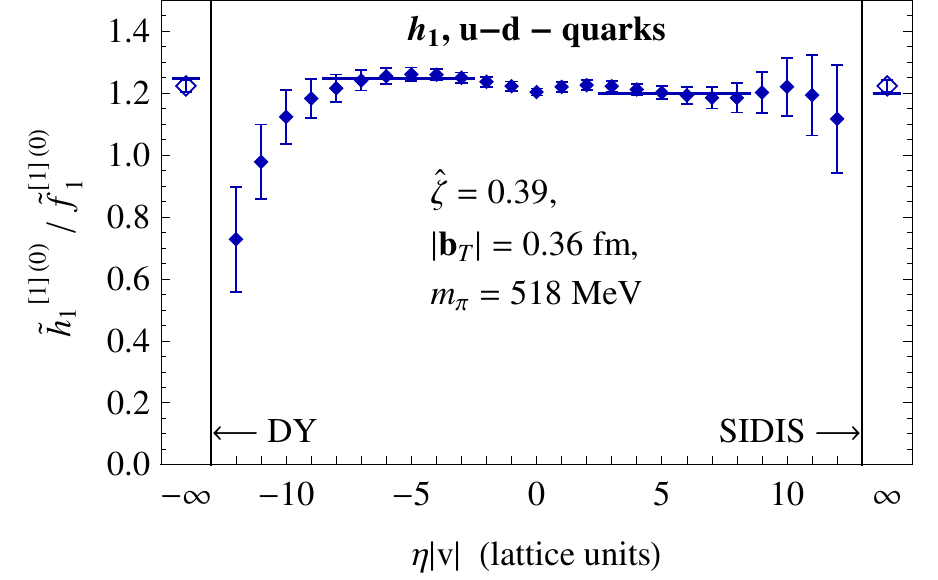}
		}\hfill%
			\subfloat[][]{%
		\label{fig-h1Rat_bdepend}%\
		\includegraphics[scale=0.9]{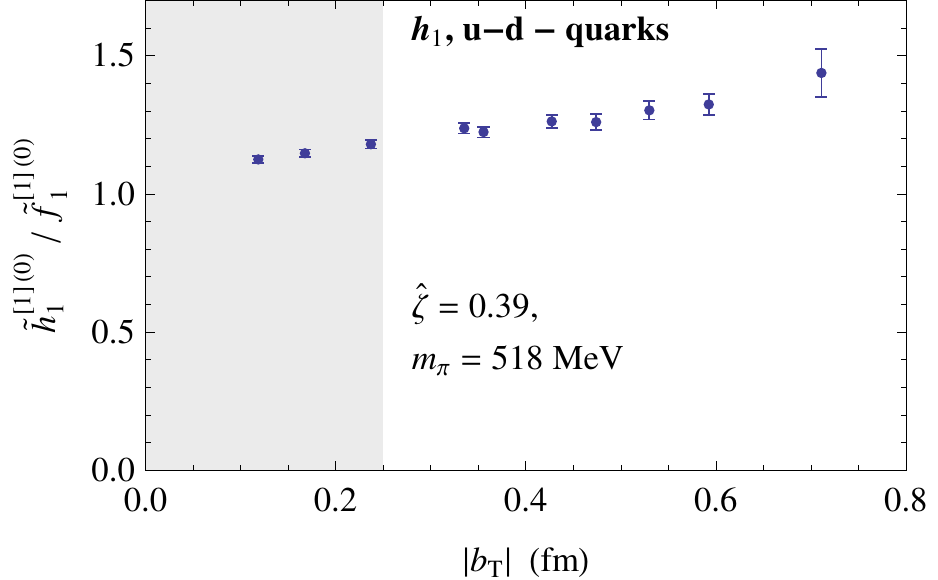}
		}\\%
	\caption[SIDIS diagram]{%
		\subref{fig-h1Rat_lsqr-1_zetasqrlat4}-\subref{fig-h1Rat_lsqr-9_zetasqrlat4} The dependence of 
		the transversity ratio $\tilde h_{1}^{[1](0)}/\tilde f_1^{[1](0)}$, Eq.~(\ref{eq-h1overf1}), on the staple extent $\eta |v|$,
		obtained at $m_\pi = 518 \units{MeV}$, $\hat \zeta = 0.39$ for three different quark separations
		$|\vprp{b}|=1a=0.12\units{fm}$, $2a=0.24\units{fm}$ and $3a=0.36\units{fm}$. 
		Asymptotic results corresponding to SIDIS and DY have been extracted as in Fig. \ref{fig-Sivers_etadepend}, except that
		we assume an even behavior of $h_{1}$ to obtain the data points plotted as empty symbols at $\eta |v| \rightarrow \pm \infty$.
		The averages (lines) are obtained from the data points with staple extents in the ranges 
		$\eta |v| = 3a .. 8a$ and $\eta |v| = -8a .. -3a$, respectively.
			Figure \subref{fig-h1Rat_lsqr-1_zetasqrlat4} might be affected by significant lattice cutoff effects due to the small quark separation
			$|\vprp{b}|=a$. \\
		\subref{fig-h1Rat_bdepend} The transversity ratio $\tilde h_{1}^{[1](0)}/\tilde f_1^{[1](0)}$ as a function of the quark separation $|\vprp{b}|$
		from the SIDIS results extracted
		on the lattice with $m_\pi = 518 \units{MeV}$ for $\hat \zeta = 0.39$.
		The data points lying in the shaded area below $|\vprp{b}| \approx 0.25 \units{fm}$ might be affected by lattice cutoff effects.
		Error bars show statistical uncertainties only.	
		%\label{fig-h1Rat_etadepend}%
		}
\end{figure}

A suitable observable for investigating these questions is the ``generalized tensor charge" given by the ratio of the (lowest $x$-moments of the)
transversity to the unpolarized distribution, $\tilde h_{1}^{[1](0)}/\tilde f_1^{[1](0)}$, defined in Eq.~(\ref{eq-h1overf1}).
The $\eta|v|$-dependence of this transversity ratio is displayed in Figs.~\ref{fig-h1Rat_lsqr-1_zetasqrlat4} to \ref{fig-h1Rat_lsqr-9_zetasqrlat4},
for different $|\vprp{b}|$ of $0.12$, $0.24$, and $0.36\units{fm}$, a pion mass of $m_\pi = 518 \units{MeV}$, and $\hat\zeta=0.39$.
We find it quite remarkable to see that $\tilde h_{1}^{[1](0)}/\tilde f_1^{[1](0)}$ stays nearly constant over the full range of accessible $|\eta||v|$ in Figs.~\ref{fig-h1Rat_lsqr-1_zetasqrlat4} and \ref{fig-h1Rat_lsqr-4_zetasqrlat4}, within comparatively small statistical errors.
For $|\vprp{b}|=0.36\units{fm}$, we see little dependence apart from larger values of $|\eta||v|$ where 
the signal-to-noise ratio quickly decreases. In all cases, we find indications for plateaus from $|\eta||v|\sim 3a \ldots 8a$.
As in the previous sections, we choose to average over the data in the plateau regions (solid lines), and 
obtain estimates for $|\eta||v|\rightarrow\pm\infty$ from the mean of the DY and SIDIS averages by imposing the symmetry condition in $\eta |v|$.
The corresponding results are illustrated by the open diamonds.
In all considered cases, differences between $|\eta||v|\rightarrow\pm\infty$ and $|\eta||v|=0$ are barely visible within uncertainties.
In other words, lattice data for simple straight gauge links provide already a very good estimate
for the phenomenologically interesting case of infinite staple extents, 
at least in the covered ranges of $\hat\zeta$ and not too large $|\vprp{b}|$.

The $|\vprp{b}|$-dependence of our estimates for the transversity ratio at $|\eta||v|=\pm\infty$ is displayed in Fig.\ref{fig-h1Rat_bdepend},
for $m_\pi = 518 \units{MeV}$ and $\hat\zeta=0.39$.
We observe a small, approximately linear rise of in total about $20\%$ as $|\vprp{b}|$ increases from $0.12 \units{fm}$ to about $0.6 \units{fm}$.
This is in agreement with our previous observation of a flatter $|\vprp{b}|$-dependence of the amplitude $\tilde A^{u-d}_{9m}$
compared to $\tilde A^{u-d}_{2}$ in Ref.~\cite{Musch:2010ka} on the basis of straight gauge links\footnote{Note again that $b$ in the present work corresponds to $-l$ in \cite{Musch:2010ka}.}.
Remarkably, a naive linear extrapolation of the data to $|\vprp{b}|=0$ would give a value for the tensor charge, 
$g_T^{u-d}=\int\! dx\, d^2\vprp{k}\, h_1(x,\vprp{k}^2)=\tilde h_{1}^{[1](0)}(\vprp{b}\!=\!0)$, of $g_T^{u-d}\approx1.1$, which is 
in very good agreement with the direct lattice calculation of this quantity using a renormalized local operator that has
been presented in \cite{Edwards:2006qx} for the same ensemble, for a scale of $\mu^2=4 \units{GeV}^2$
in the $\MSbar$ scheme\footnote{Note that the tensor charge is denoted by $\langle 1\rangle_{\delta q}$ in \cite{Edwards:2006qx}.}.

\begin{figure}[btp]
	\centering%
	\subfloat[][]{%
		\label{fig-h1Rat_evolution}%
		\includegraphics[scale=0.9]{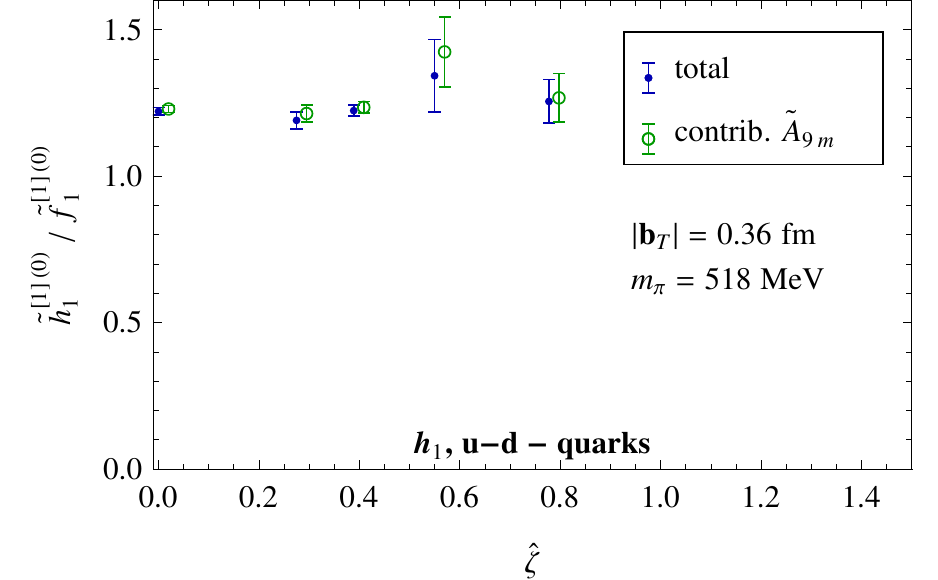}
		}\hfill%
	\subfloat[][]{%
		\label{fig-h1Rat_evolution_combined}%\
		\includegraphics[scale=0.9]{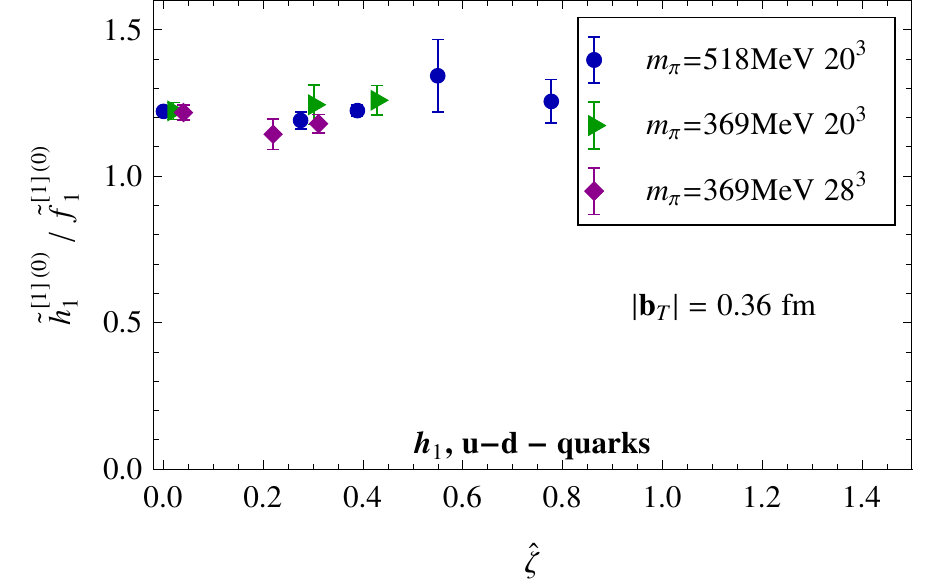}
		}\\%
	\caption[SIDIS diagram]{%
		Evolution with respect to $\hat \zeta$ for the transversity ratio $\tilde h_{1}^{[1](0)}/\tilde f_1^{[1](0)}$ at a quark separation of $|\vprp{b}|=3a=0.36\units{fm}$.
		Figure \ref{fig-h1Rat_evolution} shows the SIDIS results obtained at $m_\pi = 518\units{MeV}$. The solid data points correspond to the full result,
		and empty symbols to the result obtained with just $\tAmp_{9m}$ in the numerator. 
		Figure \ref{fig-h1Rat_evolution_combined} displays the full results for all ensembles listed in Table \ref{tab-gaugeconfs}.
		\label{fig-h1Rat_aevolution}%
		}

\end{figure}

Figure \ref{fig-h1Rat_evolution} shows the transversity ratio as a function of the Collins-Soper parameter, for the same pion mass as before but a fixed $|\vprp{b}|=0.36\units{fm}$.
The $\hat\zeta$-dependence turns out to be rather flat over the full range of accessible values.
It is interesting to note that, in contrast to the T-odd distributions discussed before, 
the amplitude $\tilde A_{9m}$ (open circles) provides $\sim100\%$ of the total results, while
the contribution from  $R(\zetahat^2) \tBmp_{15}$, Eq.~(\ref{eq-ABamps}), as well as from $\tBmp_{17}$ through Eq.~(\ref{eq-A9m}),
is negligible within errors, over the full range of $\hat\zeta$.

Finally, we show a comparison of our results for $(\tilde h_{1}^{[1](0)}/\tilde f_1^{[1](0)})^{u-d}$ obtained for the different ensembles in Fig.~\ref{fig-h1Rat_evolution_combined}. 
As before, the data points for the two values of the pion mass and the different volumes agree within uncertainties.
On the basis of the comparatively good signal-to-noise ratio for this observable, we conclude that the $\hat\zeta$-dependence 
is in this case rather flat and very well compatible with a constant behavior, 
$(\tilde h_{1}^{[1](0)}/\tilde f_1^{[1](0)})^{u-d}(\hat\zeta)\approx 1.2\pm0.1$, at least for $\hat\zeta\le0.8$ and the given parameters.
It would be interesting to investigate in future lattice studies whether this
constant behavior persists as one approaches larger Collins-Soper parameters.

\subsubsection{$T$-even TMDs: The generalized worm gear shift from $g_{1T}$}
\begin{figure}[btp]
	\centering%
	\subfloat[][]{%
		\label{fig-g1TRat_lsqr-1_zetasqrlat4}%
		\includegraphics[scale=0.9]{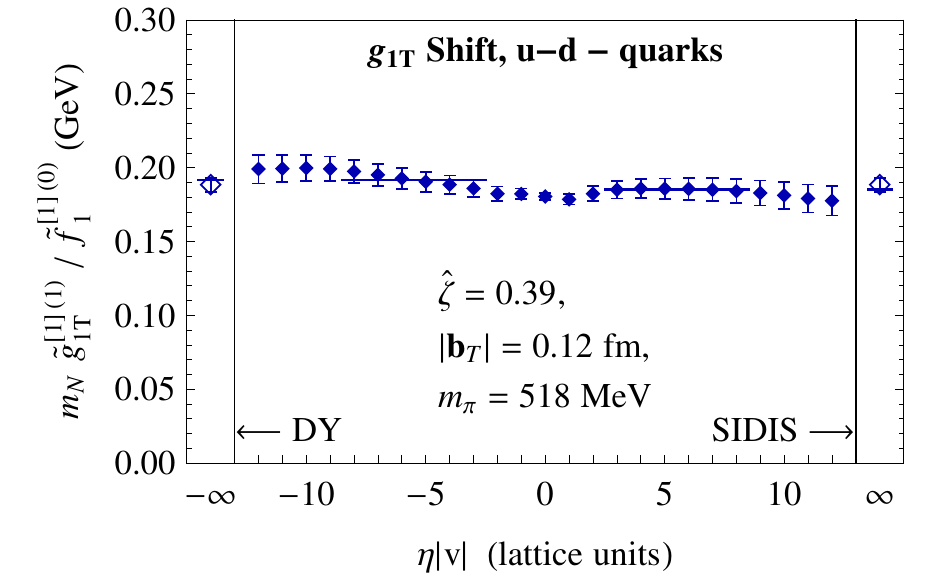}
		}\hfill%
	\subfloat[][]{%
		\label{fig-g1TRat_lsqr-9_zetasqrlat4}%\
		\includegraphics[scale=0.9]{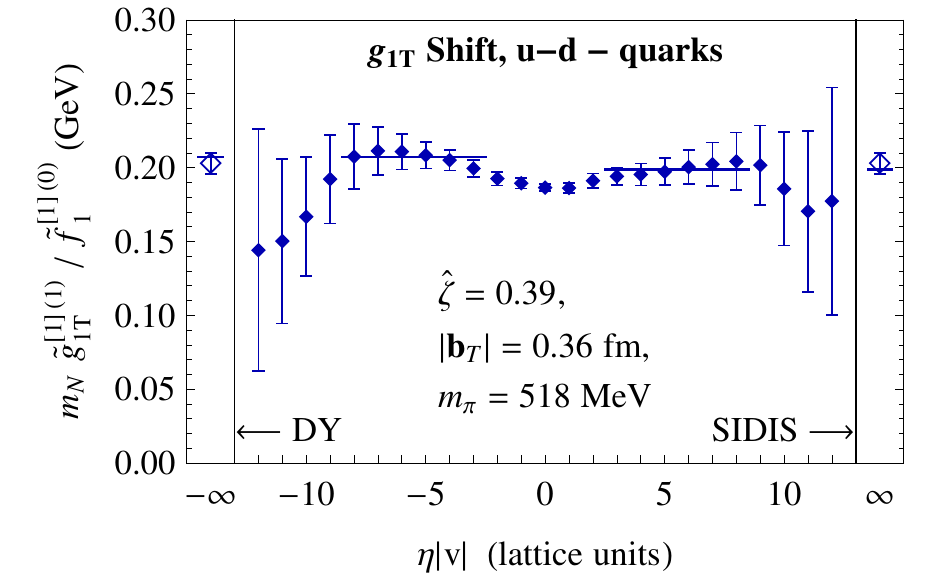}
		}\\%
	\caption[SIDIS diagram]{%
		Dependence of the generalized $g_{1T}$ shift on the staple extent $\eta |v|$,
		obtained at $m_\pi = 518 \units{MeV}$, $\hat \zeta = 0.39$ for two different quark separations
		$|\vprp{b}|=1a=0.12\units{fm}$ and $|\vprp{b}|=3a=0.36\units{fm}$. 
		Asymptotic results corresponding to SIDIS and DY have been extracted as in Figs. \ref{fig-h1Rat_lsqr-1_zetasqrlat4} to \ref{fig-h1Rat_lsqr-9_zetasqrlat4}.
		Error bars show statistical uncertainties only.			
		Figure \subref{fig-g1TRat_lsqr-1_zetasqrlat4} might be affected by significant lattice cutoff effects due to the small quark separation $|\vprp{b}|=a$.
		\label{fig-g1TRat_etadepend}%
		}
\end{figure}

As a final example, we study in this section the generalized shift defined
in Eq.~(\ref{eq-geng1TShift}), which is essentially given by the T-even
TMD $g_{1T}$. Figure~\ref{fig-g1TRat_etadepend} shows
$(\tilde g_{1T}^{[1](1)}/\tilde f_1^{[1](0)})^{u-d}$ as a function of 
$\eta |v|$ for $m_\pi = 518 \units{MeV}$, $\hat \zeta = 0.39$
and two values of $|\vprp{b}|$.
Within uncertainties, we observe, as expected, an approximate symmetry with respect to the sign of $\eta |v|$.
Furthermore, we find overall only little dependence on the staple extent for $|\vprp{b}|=0.12\units{fm}$.
At larger $|\vprp{b}|$, the signal-to-noise ratio quickly decreases as $|\eta| |v|$ becomes larger.
Still, we find indications that the results stabilize in the region $|\eta| |v| = 3a \ldots 8a$, which we choose as our plateau 
region for the computation of average values, as discussed in the previous section.

\begin{figure}[btp]
	\centering%
	\includegraphics[scale=0.9]{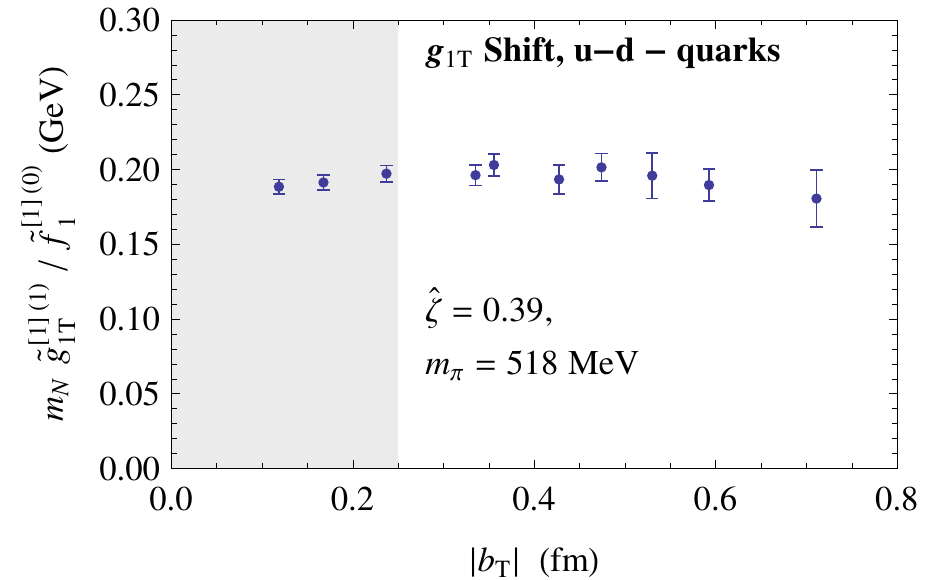}
	\caption{%
		Generalized $g_{1T}$ shift for $|\eta| |v| = \infty$ as a function of the quark separation $|\vprp{b}|$ 
		from the SIDIS and DY results extracted
		on the lattice with $m_\pi = 518 \units{MeV}$ for $\hat \zeta = 0.39$.
		The data points lying in the shaded area below $|\vprp{b}| \approx 0.25 \units{fm}$ might be affected by lattice cutoff effects.
		Error bars show statistical uncertainties only.
		\label{fig-g1TRat_bdepend}%
		}
\end{figure}

As before, the averages serve as approximations for the asymptotic results at $\eta |v| = \pm\infty$, i.e., corresponding to infinite staple extents.
The dependence of these asymptotic values on $|\vprp{b}|$ is displayed in Fig.~\ref{fig-g1TRat_bdepend}.
Although a small curvature in the central values can be observed, the results are overall rather stable within errors, with
$\langle \vect{k}_x \rangle_{u-d}^{\wg}=(\tilde g_{1T}^{[1](1)}/\tilde f_1^{[1](0)})^{u-d}\approx0.16\ldots 0.21 \units{GeV}$.

\begin{figure}[btp]
	\centering%
	\subfloat[][]{%
		\label{fig-g1TRat_evolution}%
		\includegraphics[scale=0.9]{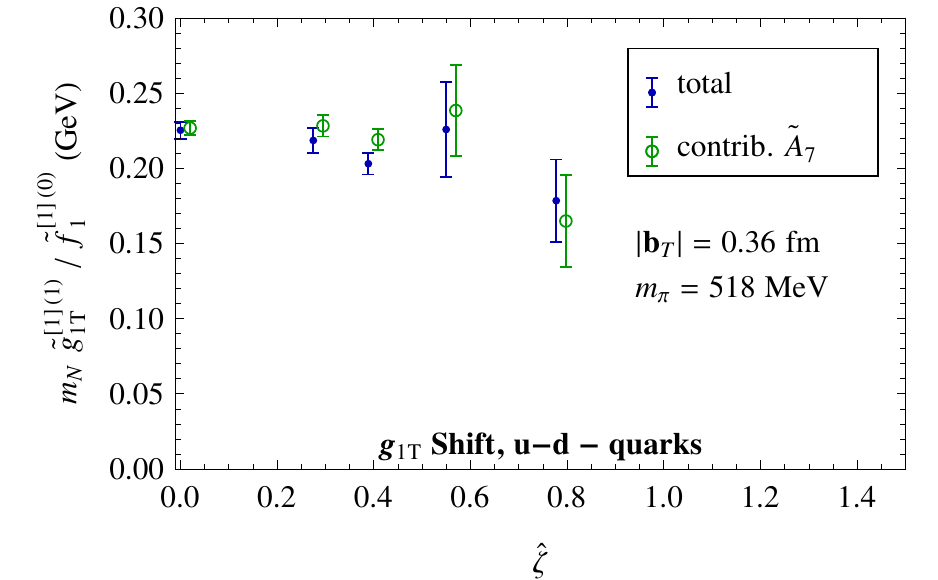}
		}\hfill%
	\subfloat[][]{%
		\label{fig-g1TRat_evolution_combined}%\
		\includegraphics[scale=0.9]{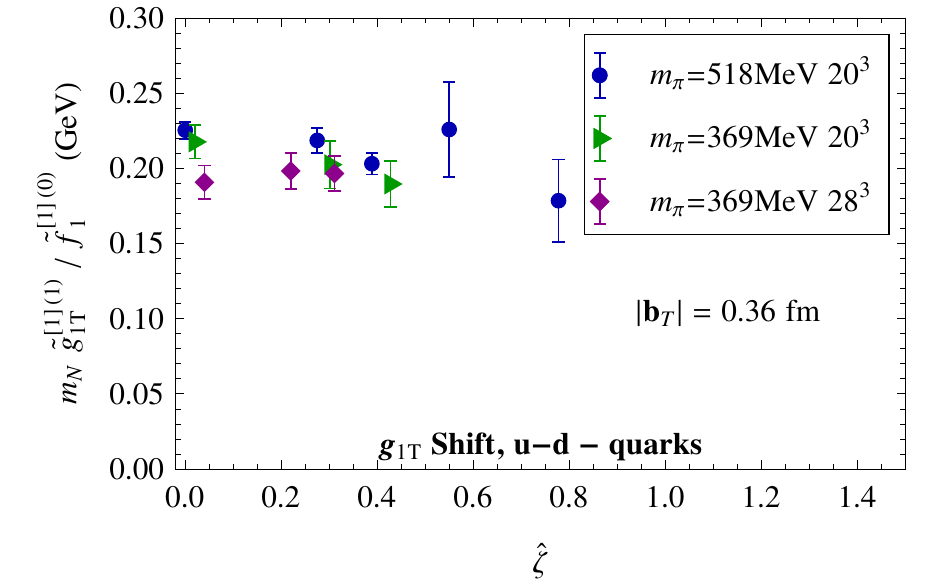}
		}\\%
	\caption[SIDIS diagram]{%
		Evolution with respect to $\hat \zeta$ for the generalized $g_{1T}$ shift at a quark separation of $|\vprp{b}|=3a=0.36\units{fm}$.
		Figure \ref{fig-g1TRat_evolution} shows the results obtained at $m_\pi = 518\units{MeV}$ for the SIDIS and DY limit $|\eta|| v|\rightarrow \infty$. The solid data points correspond to the full result
		and empty symbols to the result obtained with just $\tAmp_{7}$ in the numerator. 
		Figure \ref{fig-g1TRat_evolution_combined} displays the full results for all ensembles listed in Table \ref{tab-gaugeconfs}.
		\label{fig-g1TRat_aevolution}%
		}
\end{figure}

In Fig.~\ref{fig-g1TRat_evolution}, we show the dependence of the generalized $g_{1T}$ shift on the Collins-Soper parameter, 
for a pion mass of $m_\pi = 518 \units{MeV}$ and a fixed $|\vprp{b}|=0.36\units{fm}$.
As $\hat\zeta$ increases, one observes a slight trend towards values that are smaller in magnitude, although
it is difficult to draw any strong conclusions in view of the present uncertainties.
Similar to the case of the transversity ratio of the previous section, we find that
essentially the full signal is due to the amplitude $\tilde A_7$, while the contribution from 
$R(\zetahat^2) \tBmp_{13}$, cf. Eq.~(\ref{eq-ABamps}), is compatible with zero within errors.

Finally, Fig.~\ref{fig-g1TRat_evolution_combined} gives an overview of our results as functions of $\hat\zeta$, obtained
for the three considered ensembles, for $|\vprp{b}|=0.36\units{fm}$.
Apart from $\hat\zeta=0$, the data points for the two pion masses and volumes clearly overlap within uncertainties.
Taking into consideration the results for $m_\pi = 369 \units{MeV}$ and a spatial volume of $28^3$ (given by the filled diamonds),
the data are overall compatible with a constant behavior, although more statistics is necessary to establish a clear trend in $\hat\zeta$. 
Altogether, we observe a sizeable positive generalized transverse shift in the range of 
$\langle \vect{k}_x \rangle_{u-d}^{\wg}=(\tilde g_{1T}^{[1](1)}/\tilde f_1^{[1](0)})^{u-d}\approx0.15\ldots\ 0.25 \units{GeV}$,
for $\hat\zeta=0\ldots0.8$ and the given parameters.
We note that these values are in good agreement with our previous analyses on the basis of straight gauge links \cite{Hagler:2009mb,Musch:2010ka}.

\section{Summary and Conclusions}

\label{sec-summary}

We have presented an exploratory study of quark transverse momentum
distributions in the nucleon in full lattice QCD employing non-local operators 
with staple-shaped gauge links (Wilson lines).
Compared to our earlier works \cite{Hagler:2009mb,Musch:2010ka}, the use of staple-shaped instead of straight link paths allowed us for the first time to systematically access the naively time-reversal odd (T-odd) observables, in particular the amplitudes related to the Sivers and the Boer-Mulders TMDs.
In the framework of QCD factorization theorems, the path dependence
corresponds to a process dependence that leads to the famous sign difference
between the T-odd TMDs for the SIDIS and the DY processes. In our study, we
were able to distinguish the SIDIS and DY cases through the relative
orientation of the nucleon momentum $P$ and the vector $\eta v$ that
characterizes the direction and extent of the staple on the lattice, 
cf. Fig.~\ref{fig-link-staple}. It is important to keep in mind that TMDs
defined with non-light-like staple vectors $v$, as required on the lattice, 
will additionally depend on the Collins-Soper evolution parameter, here denoted by $\hat\zeta$.
In order to avoid additional soft factors in the formal definition of the TMDs, we have concentrated  
on the Sivers, Boer-Mulders, and worm-gear ($g_{1T}$) generalized transverse momentum shifts and the generalized tensor charge. 
Since the generalized shifts and tensor charge are defined in terms of ratios of TMDs, potential soft factors 
as well as the renormalization constants cancel out.
Our numerical results, obtained for three different ensembles with pion masses
$m_\pi = 369 \units{MeV}$ and $m_\pi = 518 \units{MeV}$, as well as spatial
lattice volumes of $20^3$ and $28^3$, are very promising: We find clearly
non-zero, sizeable signals for all observables we considered. The expected
anti-symmetry (change of sign) for T-odd quantities in $\eta |v|$ is
fulfilled within statistical uncertainties. In contrast, for the T-even
quantities we observe little systematic dependence on the staple direction
and extent. As the staple extents are increased, our data appear to approach
plateaus. Averages of the plateau values then provide estimates for the limit
of infinite staple extents, $\eta |v|\rightarrow\pm\infty$, which is formally
required for all phenomenologically relevant TMDs.

The physical length scale beyond which the influence of the gauge link
extent diminishes is of the order of $0.4\units{fm}$ for all cases considered. This observation
invites  speculations as to the physical background of this scale, e.g., an interpretation as color correlation length.
The scale might also be related to a mass gap in the spectrum. 
If $v$ is interpreted as the Euclidean time direction, the legs of the staple-shaped gauge link resemble static quark propagators. 
Considering our three-point function in this rotated frame of reference suggests that the plateau region is reached when the propagation time $|\eta v|$ of the static quark pair is large enough to suppress contributions from excited states sufficiently.

Our numerical extrapolations to infinite staple extents,
$\eta |v|\rightarrow\pm\infty$, represent first predictions for the signs and 
approximate sizes of the generalized transverse shifts from lattice QCD.
In particular, we find strong indications that the T-odd Sivers and
Boer-Mulders TMDs are both sizeable and negative for the isovector,
$u-d$ quark combination in the case of SIDIS.

Within statistical errors, we do not observe any clear trend in the data for
the transverse shifts as functions of the Collins-Soper evolution parameter
$\hat\zeta$ in the range $\hat\zeta\sim0 \ldots 0.8$.
For the T-even generalized tensor charge, which shows a much better signal-to-noise ratio and less scatter of the data points, 
we can tentatively conclude that it is approximately constant in $\hat\zeta$ for the accessible parameter ranges.
We stress, however, that more quantitative predictions with respect to phenomenological analyses 
of SIDIS and DY experiments on the basis of QCD factorization will require much larger Collins-Soper parameters $\hat\zeta \gg 1$.
For the TMD ratios discussed in this study, large $\hat\zeta$ can in principle be accessed through larger nucleon momenta.
In practice, this represents a considerable challenge due to quickly decreasing signal-to-noise ratios and potentially 
significant finite volume effects at higher $P$.
Still, we expect that future lattice results for an extended range of momenta
and with improved statistics will be very useful to establish trends in
$\hat \zeta$, eventually allowing extrapolation into the region where
factorization theorems and related evolution equations are applicable.

\begin{acknowledgments}

We thank Harut Avakian, Gunnar Bali, Alexei Bazavov, Vladimir Braun, Markus Diehl, Robert Edwards, Meinulf G\"ockeler, Barbara Pasquini, Alexei Prokudin, David Richards and Dru Renner for helpful discussions and suggestions. 
We are grateful to the LHP collaboration for providing their lattice quark
propagators to us, and for technical advice, as well as to the MILC
collaboration for use of their Asqtad configurations.
Our calculations, which relied on the Chroma software suite
\cite{Edwards:2004sx}, employed computing resources provided by
the U.S.~Department of Energy through USQCD at Jefferson Lab.
The authors acknowledge support by the Heisenberg-Fellowship program of the DFG 
(Ph.H.), SFB/TRR-55 (A.S.) and the U.S.~Department of Energy under grants
DE-FG02-96ER40965 (M.E.) and DE-FG02-94ER40818 (J.N.). M.E.~furthermore is
grateful to the Jefferson Lab Theory Center for its generous support and
hospitality during Fall 2011, which proved invaluable for the progress of
this project. Authored by Jefferson Science Associates, LLC under
U.S. DOE Contract No. DE-AC05-06OR23177. 
The U.S. Government retains a non-exclusive, paid-up, irrevocable, world-wide license to publish or reproduce this manuscript for U.S. Government purposes.
\end{acknowledgments}

%%%%%%%%%%%%%%%%%%%%%%%%%%%%%%%%%%%%%%%%%%%%%%%%%%%%%%%%%%%%%%%%%

% \pagebreak[4]

\appendix

\section{Conventions and definitions}
\label{sec-conv}

Whenever the four-vector $\bvec$ fulfills $\bvec^2 \leq 0$, we shall make use of the abbreviation $|\bvec| \equiv \sqrt{-\bvec^2}$.

In the continuum, a ``gauge link'' or ``Wilson line'' is given by the path-ordered exponential
\begin{align}
	\WlineC{\mathcal{C}_\bvec} \ & \equiv\ \mathcal{P}\ \exp\left( -ig \int_{\mathcal{C}_\bvec} d \xi^\mu\ A_\mu(\xi) \right) \nonumber \\
	& =  \mathcal{P}\ \exp\left( -ig \int_0^1 d\lambda\ A\!\left(\mathcal{C}_\bvec(\lambda)\right) \cdot \dot{\mathcal{C}}_\bvec(\lambda) \right)\, .
	\label{eq-wlinecont}
\end{align}
Here the path is specified by a continuous, piecewise differentiable function $\mathcal{C}_\bvec$ with derivative $\dot{\mathcal{C}}_\bvec$ and with $\mathcal{C}_\bvec(0)=0$, $\mathcal{C}_\bvec(1)=\bvec$.
Straight Wilson lines between two points $x$ and $y$ shall be denoted $\Wline{x,y}$ and concatenations of several straight Wilson lines (i.e., polygons) $\Wline{x,y} \Wline{y,z} \cdots$ shall be abbreviated $\Wline{x,y,z,\ldots}$.

For an arbitrary four-vector $w$, we introduce light cone coordinates $w^+ = (w^0 + w^3)/\sqrt{2}$, $w^- = (w^0 - w^3)/\sqrt{2}$ and the transverse projection $w_\prp = (0,w^1, w^2,0)$, which can also be represented as a Euclidean two-component vector $\vprp{w}=(\vect{w}_1,\vect{w}_2)\equiv(w^1,w^2)$, $\vprp{w}\tcdot\vprp{w} \geq 0$. The basis vectors corresponding to the $+$ and $-$ components shall be denoted $\nplus$ and $\nminus$, respectively, and fulfill $\nplus \cdot \nminus = 1$. 
The nucleon moving in $z$-direction has momentum $P =  P^+ \nplus + (m_N^2/2 P^+) \nminus$ and spin $S =  \Lambda (P^+/ m_N) \nplus -  \Lambda (m_N/2 P^+) \nminus + S_\prp$, $S^2 = -1$. We use the convention $\epsilon^{0123}=1$ for the totally antisymmetric Levi-Civita symbol, and introduce $\myeps_{i j} \equiv \epsilon^{-+ij}$ such that $\myeps_{1 2} = 1$.

\newcommand{\slfrac}[2]{\left.#1\middle/#2\right.}

\section{Symmetry transformation properties of the correlator}
\label{sec-symtraf}

The symmetry transformation properties of $\Phi$ used in Refs.~\cite{Ralston:1979ys,Tangerman:1994eh} need to be generalized to arbitrary link directions $v$ to arrive at the parametrization of Ref.~\cite{Goeke:2005hb}. The transformation properties of the corresponding $b$-dependent correlator $\tilde \Phi$ with the gauge link \eqref{eq-staplelink} have already been discussed in Ref.~\cite{Musch:2010ka} and are restated here for completeness:

\begin{align}
	\widetilde \Phi_\unsub^{[\GammaOp]}(\bvec,P,S,\eta v)  
	&= \widetilde \Phi_\unsub^{[\Lambda_{\slfrac{1}{2}}^{-1}\GammaOp\Lambda_{\slfrac{1}{2}}^{\phantom{-1}}]}(\Lambda \bvec,\Lambda P, \Lambda S,\eta \Lambda v)  \ ,\label{eq-lortrans} \displaybreak[0] \\
	\widetilde \Phi_\unsub^{[\GammaOp]}(\bvec,P,S,\eta v)  
	&= \widetilde \Phi_\unsub^{[\gamma^0\GammaOp\gamma^0]}(\overline{\bvec},\overline{P},-\overline{S},\eta \overline{v})  \ ,\label{eq-conpar} \displaybreak[0] \\
	\left[ \widetilde \Phi_\unsub^{[\GammaOp]}(\bvec,P,S,\eta v)  \right]^* 
	&= \widetilde \Phi_\unsub^{[\gamma^1 \gamma^3 \GammaOp^* \gamma^3 \gamma^1]}(-\overline{\bvec},\overline{P},\overline{S},-\eta \overline{v})  \ , \label{eq-contime} \displaybreak[0] \\
	\left[  \widetilde \Phi_\unsub^{[\GammaOp]}(\bvec,P,S,\eta v)  \right]^* 
	&= \widetilde \Phi_\unsub^{[\gamma^0 \GammaOp^\dagger \gamma^0]}(-\bvec,P,S,\eta v)  \ .\label{eq-conherm}
	\end{align}	

In the equations above, the matrices $\Lambda$ and $\Lambda^{\phantom{-1}}_{\slfrac{1}{2}}$ describe Lorentz transformations of vectors $x^\mu \rightarrow {\Lambda^\mu}_\nu x^\nu$ and spinors $\psi \rightarrow \Lambda^{\phantom{-1}}_{\slfrac{1}{2}} \psi$, respectively. For any Minkowski vector $w=(w^0,\vect{w})$, the space-inverted vector is defined as $\overline{w}\equiv(w^0,-\vect{w})$ .

\FloatBarrier

\bibliography{TMDs}

\end{document}